\pdfoutput=1
\documentclass[manuscript]{aastex63}

\begin{document}

\newcommand{\kms}{~km$\,$s$^{-1}$}
\def\hb1#1{\hbox to 1em{#1\hfil}}
\def\hba#1{\hbox to 3.2em{\hfil #1\hfil}}
\def\hbb#1{\hbox to 4em{#1\hfil}}
\def\hbt#1{\hbox to 15em{\hfil #1\hfil}}
\title{Emission-Line Datacubes of the HH 32 Stellar Jet}

\correspondingauthor{Patrick Hartigan}
\email{hartigan@sparky.rice.edu}

\author[0000-0002-5380-549X]{Patrick Hartigan}
\affiliation{Physics and Astronomy Dept., Rice University, 6100 S. Main, Houston, TX 77005-1892}

\author{Lynne A. Hillenbrand}
\affiliation{Department of Astronomy, MC 249-17, California Institute of Technology, Pasadena, California 91125}

\author{Matuesz Matuszewski}
\affiliation{Cahill Center for Astrophysics, California Institute of Technology,
1216 East California Boulevard, Pasadena, California 91125}

\author{Arlindo Chan Borges}
\affiliation{Department of Astronomy, MC 249-17, California Institute of Technology, Pasadena, California, 91125}

\author[0000-0002-0466-1119]{James D. Neill}
\affiliation{Cahill Center for Astrophysics, California Institute of Technology,
1216 East California Boulevard, Pasadena, California 91125}

\author{D. Christopher Martin}
\affiliation{Cahill Center for Astrophysics, California Institute of Technology,
1216 East California Boulevard, Pasadena, California 91125}

\author{Patrick Morrissey}
\affiliation{Cahill Center for Astrophysics, California Institute of Technology,
1216 East California Boulevard, Pasadena, California 91125}

\author[0000-0002-2894-6936]{Anna M. Moore}
\affiliation{Research School of Astronomy and Astrophysics, Australian National University,
Canberra, ACT 2611, Australia}

\begin{abstract}

We analyze datacubes of over 60 emission lines in the HH~32 stellar jet acquired with
the Keck Cosmic Web Imager (KCWI).  The data cover the less-explored
blue portion of the spectrum between 3586\AA\ and 6351\AA , and
have both high spectral (R $\sim$ 10000) and spatial ($\lesssim$ 1\arcsec ) resolution.
The study includes all three major ionization states of oxygen, three Balmer lines,
multiple lines of Fe~II and Fe~III, as well as the first datacubes ever acquired
for important unblended diagnostic lines such as He~II $\lambda$4686,
Ca~I $\lambda$3933, and Mg~I] $\lambda$4571.
The data cubes generally sort according to excitation, and have a relatively
continuous progression from the highest-excitation ions (He~II, O~III) through
the intermediate-excitation ions (O~I and H~I) to the lowest-excitation ions (Ca~II and Mg~I).
Merging the KCWI cubes with HST images leads to several new insights about the flow, 
including evidence for bow shocks, partial bow shocks, spur shocks, Mach
disks, jet deflection shocks, a wiggling jet, and potential shock precursors.
The most surprising result is that one of the velocity components of Fe~II
in the Mach disk suddenly increases in flux relative to other lines by a factor of two,
implying that the Mach disk vaporizes dust in the jet. Hence, jets must
accelerate or entrain dust to speeds of over 300\kms\ without destroying the grains.

\end{abstract}

\keywords{Herbig-Haro objects (722), Stellar Jets (1607),
Shocks (2086), Astrophysical dust processes (99)}

\section{Introduction} \label{sec:intro}

Since their discovery in the 1950's as nebulous emission-line objects
in the vicinity of dark clouds, HH-objects have become a primary tool
for investigating outflows associated with star formation.
The observed radial velocities and similarity to supernova remanant spectra
led \cite{schwartz75} to identify HH objects as radiatively-cooled zones in shock fronts,
and deep emission-line images revealed that these shocks generally occur within
highly-collimated jets driven from young stars \citep[e.g.][]{mf83}.
Typically denser than their surroundings, stellar jets can 
alter the morphologies of their surroundings radically as they penetrate
large distances into their ambient clouds \citep[e.g.][]{McG04}.
Together with radiation from the star, jets provide a means for
young stars to energize clouds against gravitational collapse, and 
significantly reduce the masses of the resultant stars in some simulations
\citep{hansen12}. Most current models launch stellar jets from 
magnetized accretion disks, and jets provide one of the
only ways to study this phenomenon. These flows remove
angular momentum from the disks \citep[e.g.][]{nolan17},
with implications for the formation
of gaps in disks and planetary migration at the earliest  times.

Beacuse HH jets radiate emission lines and are optically thin, they
are especially well-suited to the standard non-LTE analyses of
interstellar medium physics. For example, one can measure electron densities, 
temperatures, and ionization states from the observed emission line
ratios, though these quantities are averages over the entire emitting
region being observed. Most HH objects are resolved spatially, and with the subarcsecond
resolution afforded by HST images it is possible to resolve the
cooling distances behind individual shock waves in many cases. Velocities are high
enough to enable proper motion studies for each emission feature
across an entire flow with a temporal baseline of $\lesssim$ 10 years,
and with radial velocities as broad as several
hundred\kms\ in some objects one can perform emission-line analysis 
both spatially and as a function of velocity (i.e., within a data `cube' with
two spatial dimensions and a third dimension in velocity).
When combined with proper motion data, the only unmeasured coordinate in phase space
for these flows is the line-of-sight distance, and even that can be inferred to some
degree for simple geometries. 

This wealth of quantitative data has led to several insights concerning
jets and the sources that drive them \citep[see][for a recent review]{frank14}.
Jets from young stars become collimated within $\sim$ 100 AU of their source, and emerge
with typical opening angles of $\sim$ 5 degrees \citep[e.g.][]{reipurth00,hm07}.
They move radially from their driving stars, but often appear to `wiggle'
as the flow from the source changes in direction or if the source itself
undergoes orbital motion \citep{mr02}.
The main cause of shock waves in jets is velocity variability, and many jets
consist of a series of nested bow shocks along the axis of the flow \citep{hartigan01,lee16}.
This geometry naturally produces faster material along the axis of the jet and slower flow
along the periphery \citep[e.g.][]{cerq15}.
Quasi-stationary X-ray knots located within a few hundred AU
of the source may be related to the jet collimation process or to boundary set
up between a stellar wind and a disk wind \citep{gunther14}.

Within jets, emission line ratios imply shock velocities typically
range from 30\kms\ to 80\kms, with higher
values present in bright bow shocks. Cooling distances measured from high-spatial
resolution observations imply magnetic fields must provide the main source of
pressure in cooling zones, and lead to measurements of Alfvenic Mach numbers of a few
\citep{hw15}.  When velocity pulses brush up against slower material along the edge of the jet, 
an oblique `spur' shock propagates into the surrounding material,
and when the jet drives a bow shock it forms a Mach disk that decelerates the shock
and becomes visible as a compact knot situated near the apex of the bow shock,
often with a distinct signature in the kinematics and emission-line ratios \citep{hartigan11}.
Occasionally a stronger shock redirects a jet if the flow encounters
a dense obstacle such as a molecular cloud \citep[e.g. HH~110;][]{rrh,raga02}, while
intersecting shocks may produce short-lived hot spots known as Mach stems \citep{hartigan16}.
These modes of shock production have inspired several recent laboratory
experiments which attempt to observe how readily a given type of shock
forms and to follow as it evolves with time \citep[e.g.][]{frank14}.

Among the first group of HH objects to be cataloged by \citet{herbig74},
HH~32 is unusual both in that it is one of the few sources that has
high-excitation lines such as [O~III] 5007 \citep{brugel81}, and in that
the brightest knots are redshifted \citep{dopita78}.  The jet emanates from
the primary star, AS~353A, in a pre-main-sequence binary pair. AS~353A has
broad, bright emission lines with classic P-Cygni type absorption features that
vary with time \citep{mss82,hms86}. The source is surrounded by molecular gas \citep{edwards83},
at a distance of $\sim$ 410~pc according to \citet{gaia2}.
The bright emission-line knots HH~32A and HH~32B possess large linewidths and exhibit
profile differences in H$\alpha$, [S~II] and [O~III] that were among
the first objects to be modeled successfully as bow shocks \citep{sbr86,hrh87}. 
The first subarcsecond resolution images of the region from HST uncovered a wealth of
filamentary structures and knots along the flow, and the observed proper motions
showed that the jet must be aligned close to the line of sight \citep{curiel97}.
Data cubes of several optical emission lines in the region around HH~32A were published
by \citet{beck04}, who constructed a bow shock model for each of the
bright knots in the cube \citep[see also][]{raga04}. \citet{davis96} also 
explained the observed low-velocity H$_2$ emission associated with
knots A and B with a model where the molecular gas was confined
to the extreme wings of a bow shock.

In this paper we use the new Keck Cosmic Web Imager (KCWI)
to acquire datacubes of the entire redshifted portion of the HH~32 outflow,
including HH~32A, HH~32B, HH~32D, and the emission between HH~32D and AS~353A.
The spectra cover the blue portion of the optical between [O~II] $\lambda\lambda$3727 and
[O~I] $\lambda$6300, and have both high spectral (R $\sim$ 10000) and spatial 
($\lesssim$ 1\arcsec) resolution. Emission lines in the
blue spectral region are fainter than standard red tracers such as H$\alpha$
and [S~II] $\lambda\lambda$6720, but the blue has many more emission lines and
can trace elements and ionization states that are inaccessible in the red.
In what follows we describe the data acquisition and reduction techniques in
Sec.~\ref{sec:data}, and present the results of the data cubes in Sec.~\ref{sec:analysis}.
Sec.~\ref{sec:discussion} combines the existing HST images with the new KCWI cubes
to construct a summary model for the entire region, including the location of the
major shock waves along the flow and a discussion of their consequences in terms
of the observed line profiles and line ratios.  We bring together our
conclusions in Sec.~\ref{sec:conclusions}.

\section{Data Acquistion and Reduction} \label{sec:data}

Observations of HH32 were obtained with the Keck Cosmic Web Imager
\citep[KCWI;][]{morrissey2018}, an integral field spectrograph mounted
on the Keck II telescope, on 14 June, 2017, during the third commissioning
run for the instrument. The small slicer with a field of view
$20\arcsec\times 8.25\arcsec$ was used along with the BM grating, giving
a resolution of $R\approx 10,000$. Three separate central wavelengths were
used for the observations: 4020 \AA\ (blue), 4700 \AA\ (green), and 5950
\AA\ (red) which gives some amount of overlap between the blue and green
cubes.  For each wavelength configuration, there were two on-target exposures
and an adjacent-in-time offset sky exposure, all of duration 600 sec.
Due to a rising moon, the blue cube was acquired first, followed by green and
then red.  The instrument position angle was set to $105^\circ$ in order to
roughly align the field with the outflow axis. Flux calibration was
achieved using nearby-in-time observations of  the spectophotometric
standard BD+25$^\circ$4655.

Data cube processing generally followed the procedures described
in Section 4 of \cite{morrissey2018}.  In summary, bias frames, internal
lamp exposures including arc lines, continuum flat and continuum bars exposures, 
and external dome flats were acquired the morning after the science observations were taken.
The arc lamp and the continuum bars exposures are used define the geometry of
the raw data, while the continuum flats and the dome flats are used to
correct illumination non-uniformities.

Once the geometry has been defined and the illumination has been corrected,
the sky exposures were used to define a sky model using basis splines that 
accurately follow the sky spectrum as a function of wavelength.  This
results in a low-noise sky model that is then subtracted from each
science image.  Offset sky observations were used since the filling
factor of object light in the on-target observations was a large fraction of 
the KCWI FOV, causing over-subtraction if the sky were modeled from the
on-target observations.  

The slicer design of KCWI means that the ends of each cube are subject to
nonuniformity  in the spatial-spectral coverage, resulting in banding in the
reduced cubes with approximately 20\% of the nominal wavelength range of the
cubes affected. Fortunately, the blue and green cubes overlap sufficiently
such that there is no loss in wavelength coverage, while there is a real gap
between the reduced green and red cubes.  For our data set, the effective total
wavelength coverage is 3586\AA $-$ 5136\AA\  and then 5540\AA $-$ 6351\AA.

Following processing, the two on-target exposures for each of the three
wavelength ranges were averaged, and the three cubes were then spatially
registered.  This involved fitting two-dimensional gaussians to the
H$\gamma$ line in the blue and green cubes, the H$\beta$
line in the green cube, and the \ion{O}{1} 6300\AA\ line in the red cube so
as to determine the peaks. The three cubes were then aligned and trimmed to
encompass the same spatial area. 

The full cubes have a native pixel asymmetry, with 0.34\arcsec\ pixels 
in the ``x" direction from the individual image slices, and
0.1457\arcsec\ pixels in the ``y" direction on the detector.  Each cube was
resampled to square pixels by binning by a factor of three in the
y-direction and regridding in x-direction to yield 0.437\arcsec\ pixels in
each dimension, which sufficiently subsamples the 0.7-1.0\arcsec\ seeing conditions
during the observations while still providing good signal-to-noise in the fainter
lines.

In most cases we want to examine how the brightness of a specific emission line 
varies spatially and with velocity, and then compare its datacube with those
of different lines. To extract a datacube for one line, we first rebin the
data from its dispersion of 0.25 \AA\ per pixel to a cube that spans $-$150\kms\
to 600\kms\ centered on the rest wavelength of the emission line in the
frame of reference of AS~353A. The LSR velocity of AS~353A is +8\kms\
from molecular line surveys \citep{edwards83}.

Although observing in the blue part of the spectrum opens up opportunities to
study many fainter lines that have been to date largely ignored, this 
innovative aspect to the data set is also in a sense a weakness in that
KCWI detects enough lines that in a number of cases the broad velocity
extent of the HH~32 outflow causes them to blend. Even though 
emission lines are optically thin in HH shocks, it is not possible to deblend lines
reliably if they arise from different elements. However, we were able to deblend
the [O~II] 3726.03\AA\ and [O~II] 3728.82\AA\ lines, separated in velocity by 
225\kms, in the following manner.  Because the electron densities inferred from the red [S~II] lines
in HH~32 \citep[e.g.][]{brugel81,beck04} are typically a factor of five higher than the critical
densities for [O~II] $\lambda$3726 and [O~II] $\lambda$3729 ($\lesssim$ 70 cm$^{-3}$), the flux ratio between these [O~II]
lines will be g$_3$A$_{31}$/g$_2$A$_{32}$, where subscripts 3, 2, and 1 refer
respectively to the $^2D_{3/2}$, $^2D_{5/2}$, and $^4S_{3/2}$ levels of [O~II],
and the 3-1 and 2-1 transitions are $\lambda$3726.03 and $\lambda$3728.82, respectively.
Using the atomic parameters compiled by \citet{mendoza83}, in the
high density limit I$_{3726}$/I$_{3729}$ = 2.88. 
Because there is no emission blueward of $-$30\kms, we can use the 3726.03\AA\ profile
to trace the velocity range of $-$60\kms\ through +195\kms. Likewise, there is
no emission redward of +450\kms, so we use the 3728.82\AA\ profile scaled by 2.88 
for velocities greater than +225\kms. At intermediate velocities the known separations
and flux ratios suffice to solve for the profile shape.

Finally, terrestrial emission lines are not always subtracted perfectly in the data
reduction procedure. We corrected for residual sky lines in the 
relevant data slices by fitting a spline to the emission along the slit, using
the pixels at top and bottom of the datacube to measure sky. This procedure worked
well for [O~I] $\lambda$6300, but failed for [O~I] $\lambda$5577, where the night sky line residuals
are too large and variable. The high-velocity portions of the [O~I] $\lambda$5577 cube are
unaffected by the residual night sky, but show no significant differences from 
the [O~I] $\lambda$6300 cubes at those velocities. Most of the data reduction beyond
the standard pipeline reductions, such as extracting data cubes of
individual lines, deblending data cubes, and improving sky subtraction were
done with various routines in IRAF.

\section{Emission Line Analysis} \label{sec:analysis}

\subsection{System Overview}

\begin{figure}
\centering
\includegraphics[angle=0,scale=1.00]{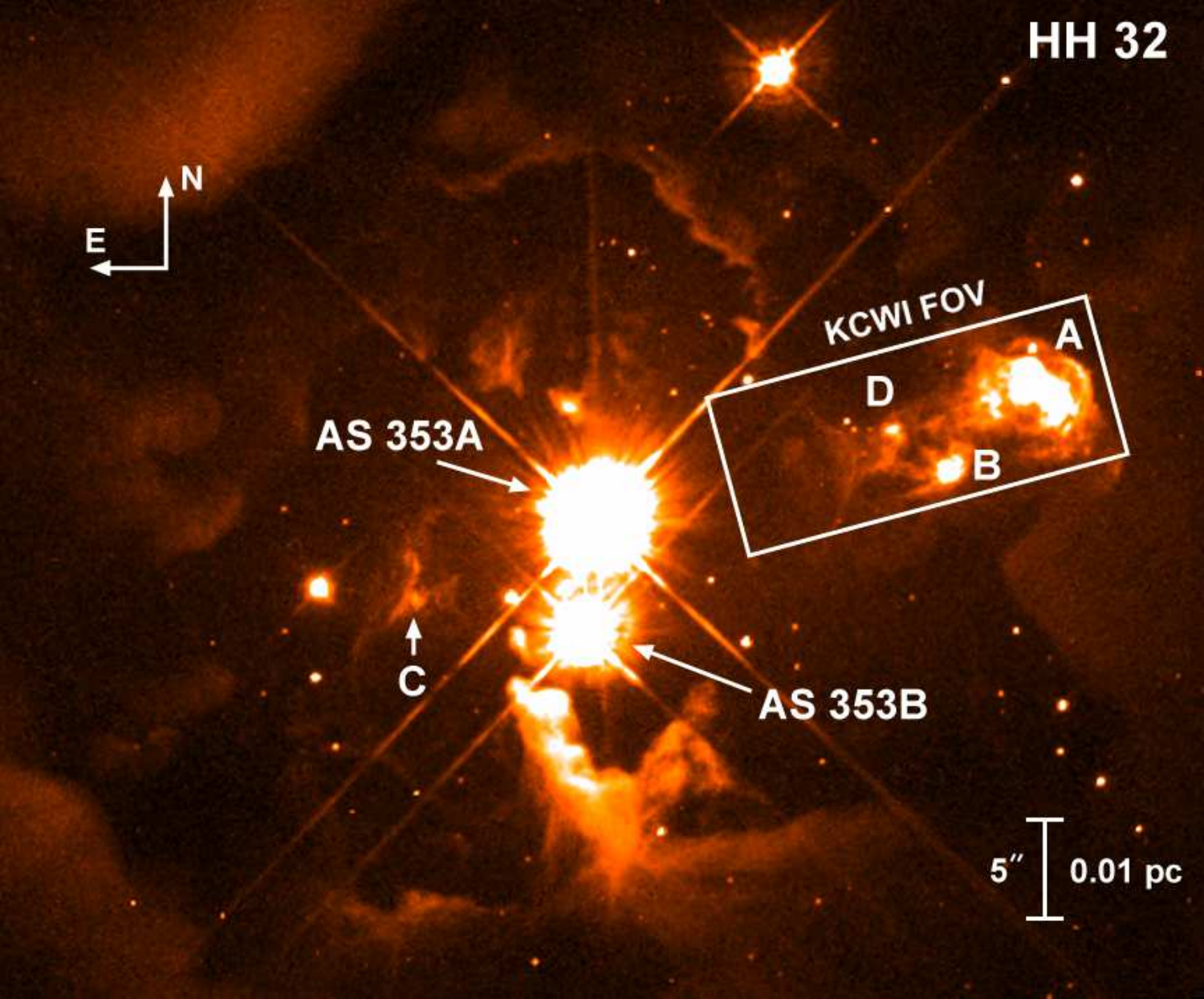}
\caption{Overview of the HH 32 Jet and its exciting star AS~353A, showing
the spatial coverage and orientation (PA = 105$^\circ$)
of the new KCWI data cubes superposed upon an archival
r-band image from HST that includes H$\alpha$ and the red [S~II] lines.
The scale bar assumes a distance of 410~pc \citep{gaia2}. The bright
jet knots A, B, and D are all redshifted, while knot C (not
observed by KCWI) is blueshifted. The nebulous structures not labeled
as part of the jet are reflection nebulae. The jet is oriented close to
the line of sight.
}
\label{fig:overview}
\end{figure}

Fig.~\ref{fig:overview} presents an overview of the HH~32 jet
and shows the extent of the KCWI images. The background r-band
(F675W) image in this figure was taken with HST on 25 Aug, 1994 as
part of program GO-5367 (PI: Raymond).  Throughout the paper we use HST images
from this program as a guide to what the structure of the jet looks like at
high spatial resolution. However, we must bear in mind that 
in the 25 years since the HST data were acquired
the jet knots have moved $\sim$ 1\arcsec $-$ 2\arcsec\ away from
AS 353A, and differential motions within the jet will shift features relative to
one another by up to an arcsecond \citep{curiel97}. The differential
motions are on the order of the ground-based
seeing and do not affect how we interpret images, though
one cannot expect a perfect correspondance between what we observe
today and what appears in the HST images from 1994.

\begin{figure}
\centering
\includegraphics[angle=0,scale=1.80]{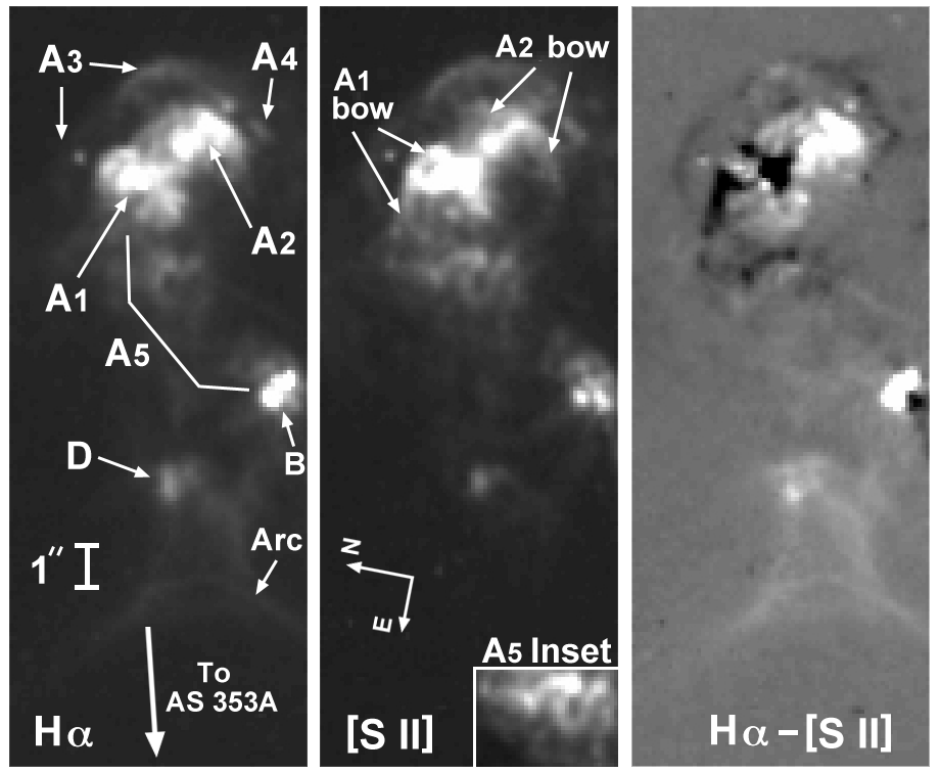}
\caption{Narrow-band HST images in H$\alpha$ (F656N; left) and [S~II] (F673N; center) 
and a difference image (right) of the HH~32 jet for the region of the KCWI observations
shown in Fig.~\ref{fig:overview}.  The scale bar of 1\arcsec corresponds to 410~AU.
The knot nomenclature follows that of \cite{beck04}. The inset of feature A5
is scaled to highlight the sinuous jet at that location.
}
\label{fig:hstdiff}
\end{figure}

The environs of AS~353A and AS~353B are quite remarkable. Reflection
nebulae in the HST image outline an intricate cavity shape
around the binary pair. Within the jet, knot A splits into two main
components, each appearing like a bow shock in the images and each
showing the broad emission and spatial separations of high- and low-velocity
emission expected from a simple bow shock model \citep{beck04}. Knot~B
occurs along the edge of the flow and also resembles a bow shock in
existing spectral maps \citep{beck04}, while knot~D has a more extended
arcuate shape visible in the Balmer lines of H.  Much structure appears
in both the H$\alpha$ and red [S~II] HST images as well as in their
difference image (Fig.~\ref{fig:hstdiff}). The framework for interpreting such
images is described in \cite{hartigan11} - filamentary Balmer lines designate the shock
fronts, and forbidden line emission such as from [S~II] follows in
a spatially-resolved cooling zone. We discuss each of the bright knots in
detail in Sec.~\ref{sec:discussion}.

\subsection{KCWI Datacubes of Emission Lines}

Our three wavelength settings detected over 60 emission lines in 
HH~32 (Tables~1 and~2).
\cite{md09} catalogued a large number of emission lines
in their comprehensive study of HH~202 in Orion, and many of
these also occur in HH~32, though we do not see the Orion H~II
recombination lines in our data.  Eliminating the faintest lines, blends, and
lines where a portion of the emission profile is lost off the end
of the CCD, we are left with 27 emission lines that have their own data cubes.
Lines from the same element and ionization state often appear
identical, so we can combine these to increase the signal-to-noise
in the final products (Table~1).

\begin{center}
\begin{deluxetable}{lll}
\tablenum{1}
\tablecolumns{3}
\tabcolsep=2em
\parindent=0em
\tablecaption{Bright Emission Lines Included in a Datacube}
\tablehead{\hspace{2em} Line\hspace{1.7em} Figure&\hspace{2em} Line\hspace{1.7em} Figure&\hspace{2em} Line\hspace{1.7em} Figure}  
\startdata
\hba{H$\beta$} \hbb{4861.33} \hb1{5}&
\hba{[O II]} \hbb{3726.03$^{c}$} \hb1{4}&
\hba{[Fe II]} \hbb{4243.97$^{d}$} \hb1{6}\\
\hba{H$\gamma$} \hbb{4340.47$^{a}$} \hb1{5} &
\hba{[O II]} \hbb{3728.82$^{c}$} \hb1{4}&
\hba{[Fe II]} \hbb{4287.39} \hb1{6}\\
\hba{H$\delta$} \hbb{4101.74} \hb1{5}&
\hba{[O III]} \hbb{4958.91} \hb1{3}&
\hba{[Fe II]} \hbb{4814.55} \hb1{6}\\
\hba{He I} \hbb{4471.47} \hb1{4}&
\hba{[O III]} \hbb{5006.84} \hb1{3}&
\hba{[Fe II]} \hbb{4889.68} \hb1{6}\\
\hba{He I} \hbb{5875.64} \hb1{4}&
\hba{[Ne III]} \hbb{3868.75} \hb1{3}&
\hba{[Fe II]} \hbb{4905.37} \hb1{6}\\
\hba{He II} \hbb{4685.70$^{b}$} \hb1{3}&
\hba{Mg I]} \hbb{4571.10} \hb1{6}&
\hba{[Fe II]} \hbb{5111.65} \hb1{6}\\
\hba{[N II]} \hbb{5754.64} \hb1{4}&
\hba{[S II]} \hbb{4068.60} \hb1{6}&
\hba{[Fe III]} \hbb{4658.10} \hb1{4}\\
\hba{[O I]} \hbb{6300.30} \hb1{5}&
\hba{[S II]} \hbb{4076.35} \hb1{6}&
\hba{[Fe III]} \hbb{4701.62} \hb1{4}\\
&
\hba{Ca II} \hbb{3933.66} \hb1{6}&
\hba{[Fe III]} \hbb{4881.00} \hb1{4}\\
\multicolumn{3}{l}{a: Observed in both green and blue cubes}\\
\noalign{\vspace {-0.1in}}
\multicolumn{3}{l}{b: Blend of 5 lines in a 26\kms\ interval}\\
\noalign{\vspace {-0.1in}}
\multicolumn{3}{l}{c: [O~II] 3726+3729 are blended, but can be combined into a single cube (see text)}\\
\noalign{\vspace {-0.1in}}
\multicolumn{3}{l}{d: Possibly blended with very weak [Fe~II] 4244.85}\\
\noalign{\vspace {0.1in}}
\enddata
\end{deluxetable}
\end{center}
                
\begin{deluxetable}{lllll}
\tablewidth{0pt}
\tabcolsep=0em
\tablenum{2}
\tablecolumns{5}
\tablecaption{Other Lines Detected}
\tablehead{\null\hspace{3.5em}&\null\hspace{3.5em}&\null\hspace{3.5em}&\null\hspace{3.5em}&\null\hspace{3.5em}}
\startdata
\noalign{\smallskip}
\multicolumn{5}{c}{\it Faint Lines, Not Included in a Datacube}\\
\hba{H-10} \hbb{3797.90} \hb1{}&\hba{[Ar IV]} \hbb{4711.37} \hb1{}&\hba{[Fe II]} \hbb{4276.83}\hb1{}&\hba{[Fe II]} \hbb{4457.95}\hb1{}&\hba{[Fe II]} \hbb{5746.97}\\
\hba{H-9} \hbb{3835.39} \hb1{}&\hba{[Ar IV]} \hbb{4740.16} \hb1{}&\hba{[Fe II]} \hbb{4346.85}\hb1{}&\hba{[Fe II]} \hbb{4728.10}\hb1{}&\hba{[Fe III]} \hbb{4754.83}\\
\hba{He I} \hbb{4921.93} \hb1{}&\hba{[S III]} \hbb{6312.10} \hb1{}&\hba{[Fe II]} \hbb{4474.91}\hb1{}&\hba{[Fe II]} \hbb{4774.75}\hb1{}&\hba{[Fe III]} \hbb{4769.90}\\
\hba{He I} \hbb{5015.68} \hb1{}&\hba{[Fe II]} \hbb{4114.48} \hb1{}&\hba{[Fe II]} \hbb{4452.10}\hb1{}&\hba{[Fe II]} \hbb{4973.42}\hb1{}&\hba{[Fe III]} \hbb{4987.20}\\
\noalign{\medskip}
\multicolumn{5}{c}{\it Blends}\\
\multicolumn{5}{c}{\hbt{H-8 3889.05 + He I 3888.65} \hba{} \hbt{[O III] 4363.21 + [Fe II] 4359.33}}\\
\multicolumn{5}{c}{\hbt{[O I] 5577.37 + Night Sky} \hba{} \hbt{[Fe II] 4416.27 + [Fe II] 4413.76}}\\
&\multicolumn{3}{c}{\hbox to 22em{Ca II (H) 3968.47 + [Ne III] 3868.75 + H-7 3970.07}}&\\
\noalign{\medskip}
\multicolumn{5}{c}{{\it Lines Detected, But Not Fully Mapped By KCWI}}\\
{[N I]} 5200.26 &{[O I]} 6363.78 & {[Fe II]} 5527.34& {[Fe II]} 5158.03& {[Fe II]} 5158.80\\
{[N I]} 5197.90&&&&\\
\enddata
\end{deluxetable}

\begin{figure}
\centering
\includegraphics[angle=0,scale=0.83]{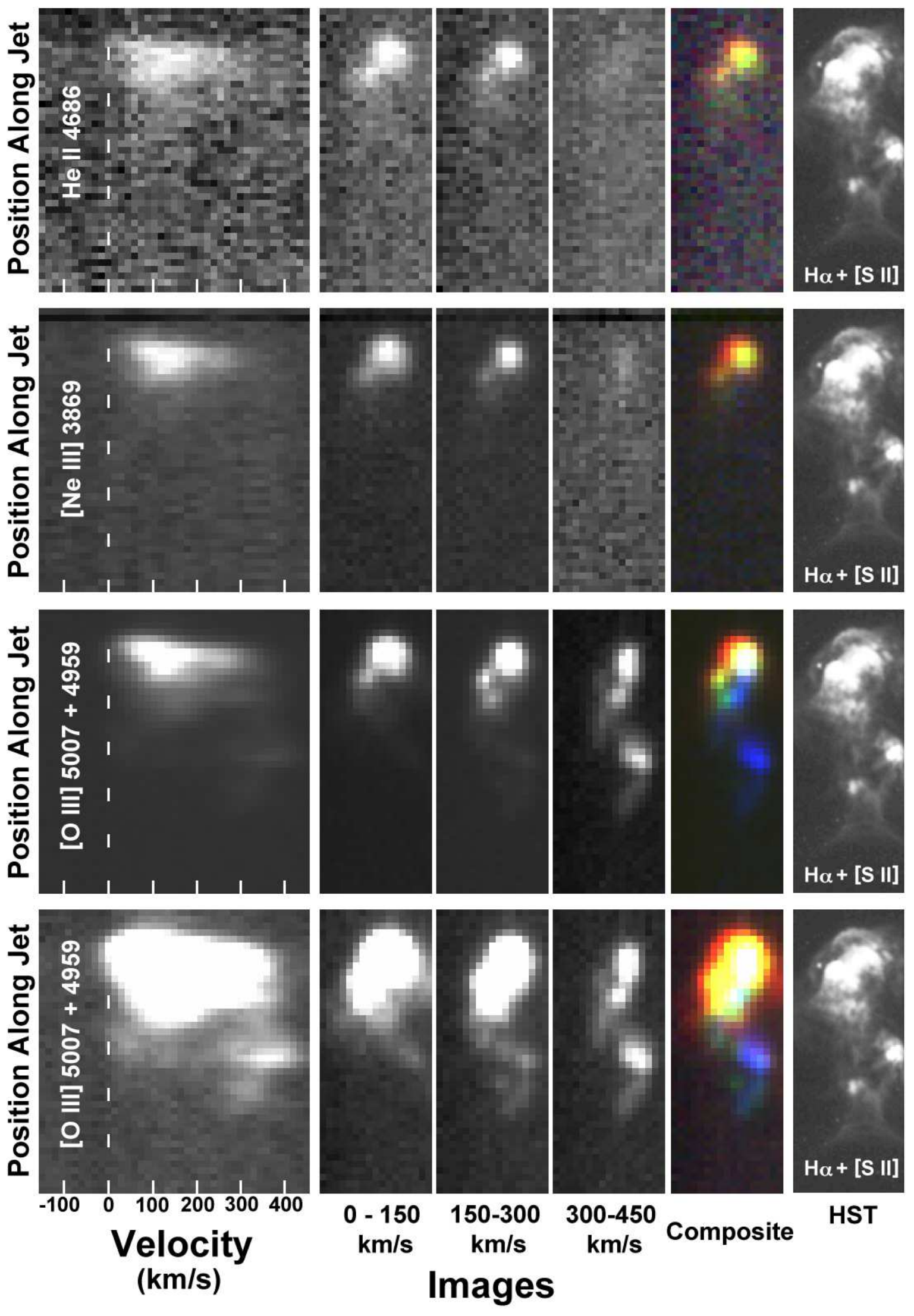}
\caption{KCWI data cubes of the highest-excitation lines: He~II 4686, [Ne~III] 3869, and [O~III]
5007+4959. Left: position-velocity
diagrams coadded across the width of the jet. The jet moves from the bottom to the top in the figure.
Center: Velocity images summed over 150\kms\
intervals. Right: Color composite of the redshifted jet, where the red, green, and blue channels
are taken from the slowest (0\kms\ $-$ 150\kms),
intermediate (150\kms\  $-$ 300\kms), and fastest (300 \kms\ $-$ 450\kms)
images, respectively. Far right: HST r-band image in Fig.~\ref{fig:overview}. Two scalings are
shown for [O~III] 5007+4959.
}
\label{fig:cube-hi}
\end{figure}

\begin{figure}
\centering
\includegraphics[angle=0,scale=0.96]{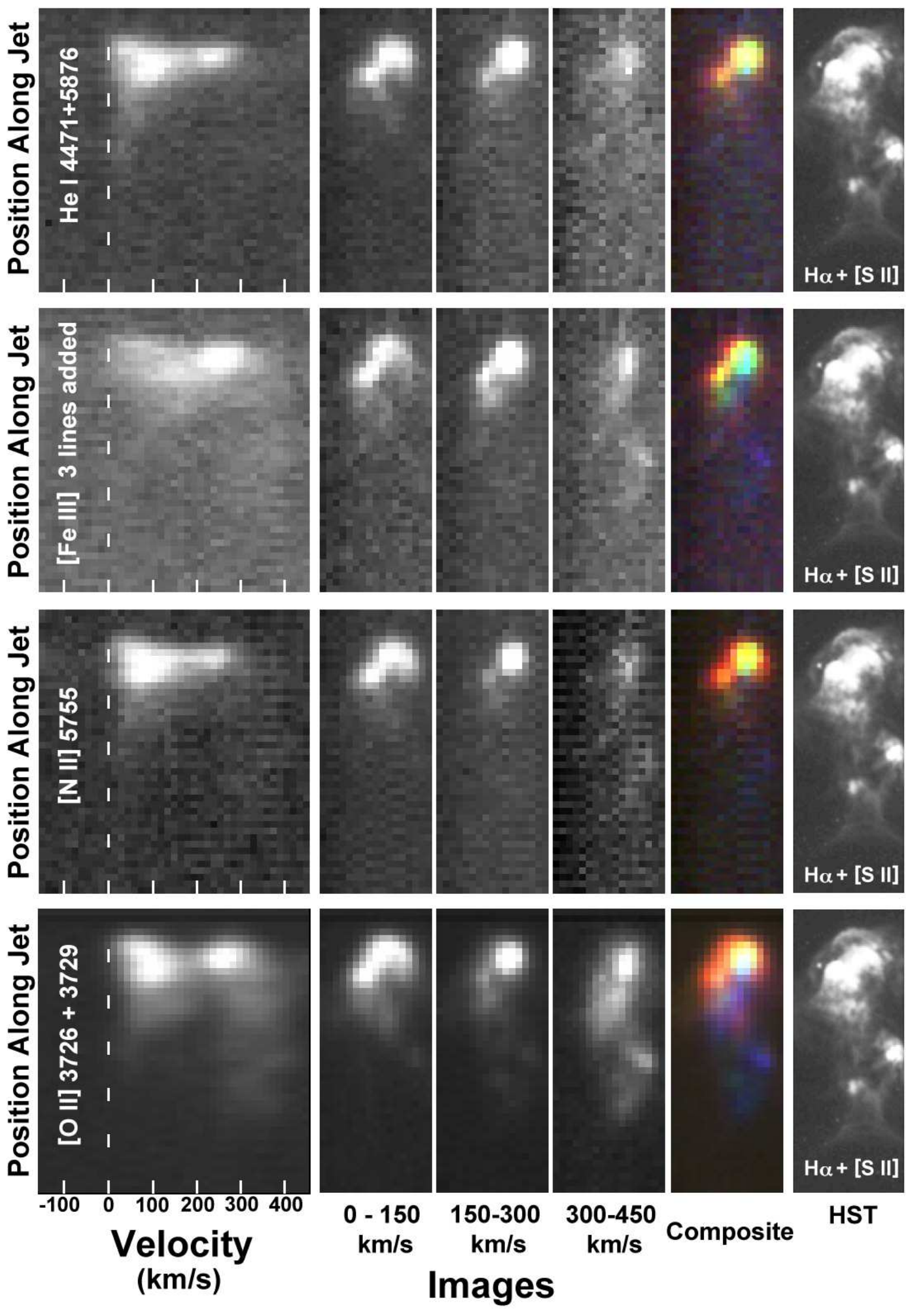}
\caption{Same as Fig.~\ref{fig:cube-hi} but for the intermediate/high-excitation lines:
He~I 4471+5876, [Fe~III] 4755+4770+4987, [N~II] 5755, and [O~II] 3726+3729.
}
\label{fig:cube-intermediate-hi}
\end{figure}

\begin{figure}
\centering
\includegraphics[angle=0,scale=0.96]{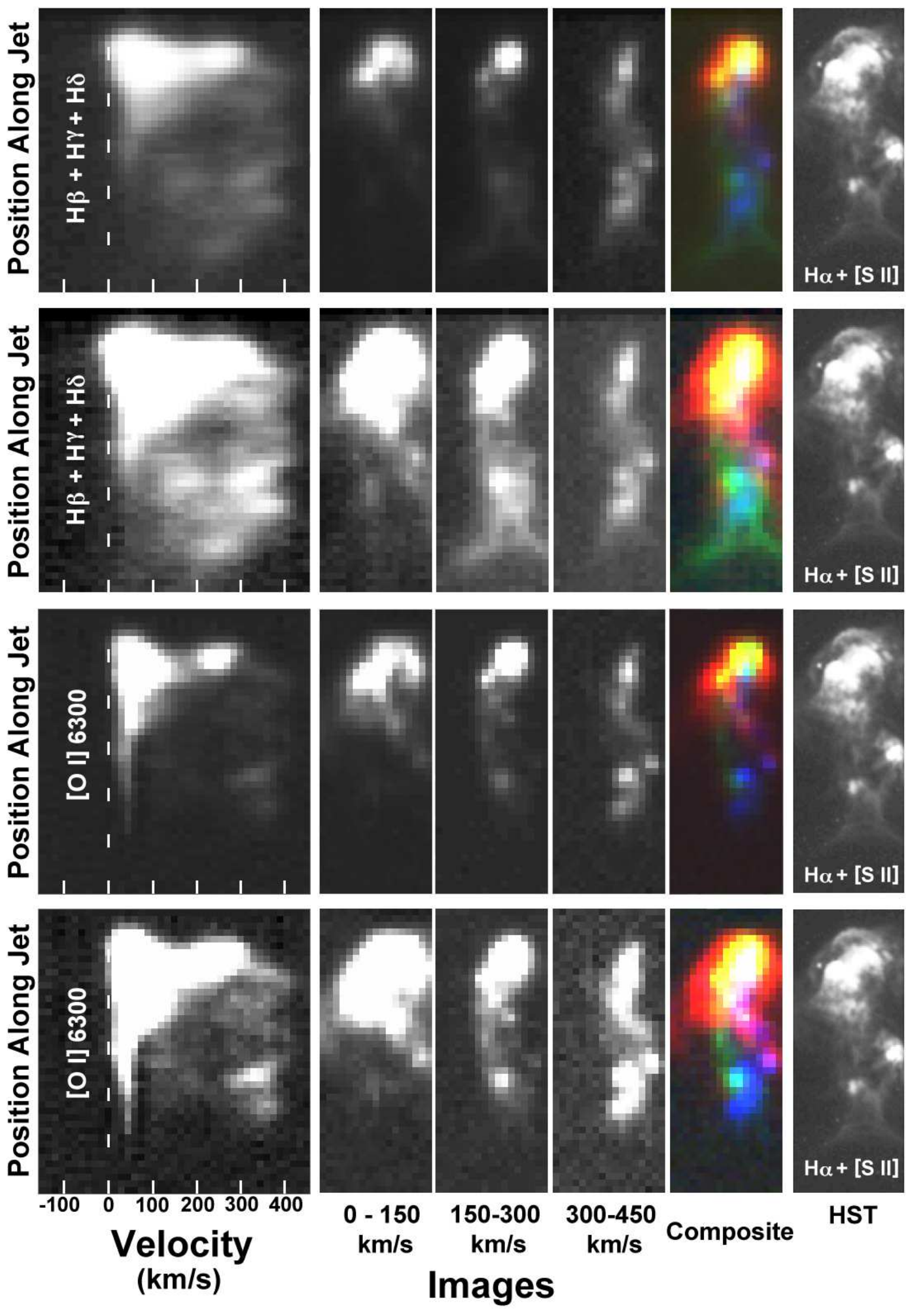}
\caption{Same as Fig.~\ref{fig:cube-hi} but for H~I (H$\beta$ + H$\gamma$ + H$\delta$)
and [O~I] $\lambda$6300. Two scalings are
shown for each line in order to highlight structure in both the brightest and faintest
parts of the diagrams.
}
\label{fig:cube-intermediate-low}
\end{figure}

\begin{figure}
\centering
\includegraphics[angle=0,scale=0.96]{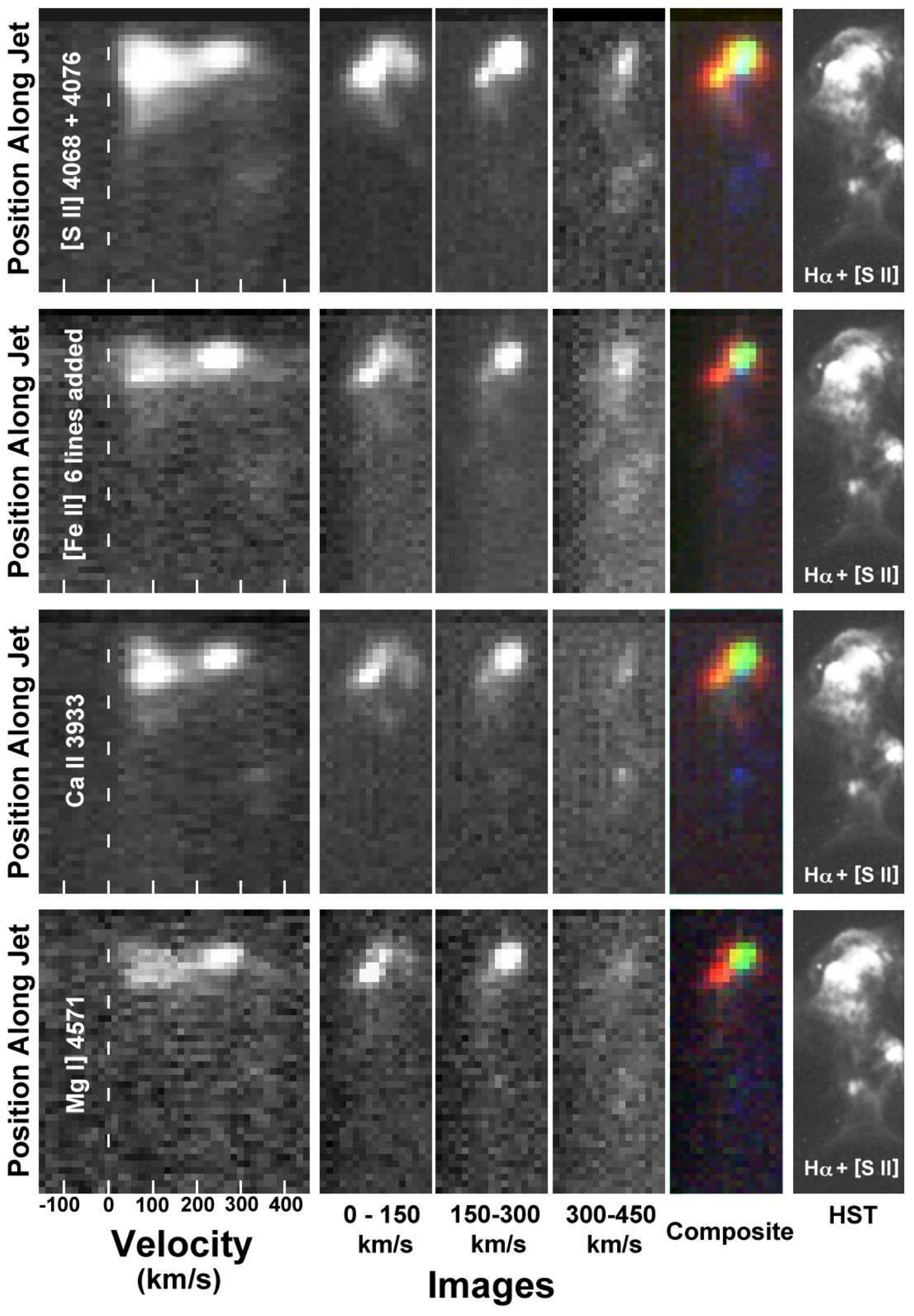}
\caption{Same as Fig.~\ref{fig:cube-hi} but for the lowest-excitation lines:
[S~II] 4068+4076, [Fe II] 4244+4287+4815+4890+4905+5112,
Ca~II (K) 3933, and Mg~I] 4571.
}
\label{fig:cube-low}
\end{figure}

We group the final data cubes
approximately according to excitation, with the highest-excitation cubes
(He~II, Ne~III and O~III) in Fig.~\ref{fig:cube-hi}, the next
group (He~I, Fe~III, N~II and O~II) in Fig.~\ref{fig:cube-intermediate-hi},
the third set (H~I and O~I) in Fig.~\ref{fig:cube-intermediate-low},
and the lowest-excitation group (S~II, Fe~II, Ca~II and Mg~I) in
Fig.~\ref{fig:cube-low}. We take `excitation' to be a sum of
the energy needed to ionize the atom to the ionization state of
interest, plus the excitation of the level above ground. For
example, He~II 4686 requires 24.6~eV to ionize He~I to He~II,
plus another 51~eV to populate the upper state of the transition.
Similarly, [O~III] 5007 needs 35.1~eV to doubly-ionize oxygen, plus
another 2.5~eV to populate the upper state. On the other hand, it
makes sense to group [O~I] 6300 with the H~I Balmer lines ($\sim$ 11~eV)
because the ionization state of oxygen is tied to that of hydrogen through
a large charge exchange coefficient \citep{williams73}.

Figs. \ref{fig:cube-hi} - \ref{fig:cube-low} show that lines with roughly the
same excitation generally have very similar-looking data cubes. It is probably
easiest to see differences in the cubes by looking carefully at the emission
from knots A1 and A2 near the top of the low-velocity (0\kms\ $-$ 150\kms) images. 
Emission in the high-excitation cubes comes mainly from the leading knot
A2, while knot A1 is the brighter of the two in the low-excitation cubes.
This difference also appears in the position-velocity diagrams. 
Two emission lines with roughly the same level of excitation ought to produce
similar data cubes in shock-excited gas, and this correspondence is indeed what we
observe in the KCWI spectral cubes.

\subsection{Reddening, Ionization and Velocity in the Jet}
\label{sec:vel}

Ratios of the H$\gamma$/H$\beta$ and H$\delta$/H$\beta$ cubes are unremarkable,
and show roughly constant values across all regions of bright emission. The observed
values of 0.31 $\pm$ 0.03 for H$\gamma$/H$\beta$ and 0.14 $\pm$ 0.015 for H$\delta$/H$\beta$
agree well with previously observed values of 0.298 and 0.162, respectively \citep{brugel81}.
Balmer line ratios in high-velocity shocks like those in HH~32 should resemble
those expected from Case B recombination \citep[][]{hrh87}, so in principle we can use
the Balmer line ratios to estimate reddening. Using a standard extinction law
and taking Balmer ratios at $10^4$~K \citep[tables~7.2 and~4.4 of][]{osterbrock89},
we estimate the logarithmic
extinction at H$\beta$, C$_{{\rm H}\beta}$ = 1.38 $\pm$ 0.30 from the H$\gamma$/H$\beta$
ratio, and C$_{{\rm H}\beta}$ = 1.41 $\pm$ 0.25 from H$\delta$/H$\beta$. For comparison,
\citet{brugel81} found C$_{{\rm H}\beta}$ = 1.05 $\pm$ 0.07 using the ratio of transauroral to auroral
[S~II] lines, a method that is superior to ours in that it spans a much larger 
wavelength range, although flux calibrations for the faint near-infrared [S~II] lines
can be challenging. These estimates all imply a rather large reddening along the line
of sight to HH~32; for example, C$_{{\rm H}\beta}$ = 1.38 implies A$_V$ $\sim$ 2.75.

\begin{figure}
\centering
\includegraphics[angle=0,scale=1.00]{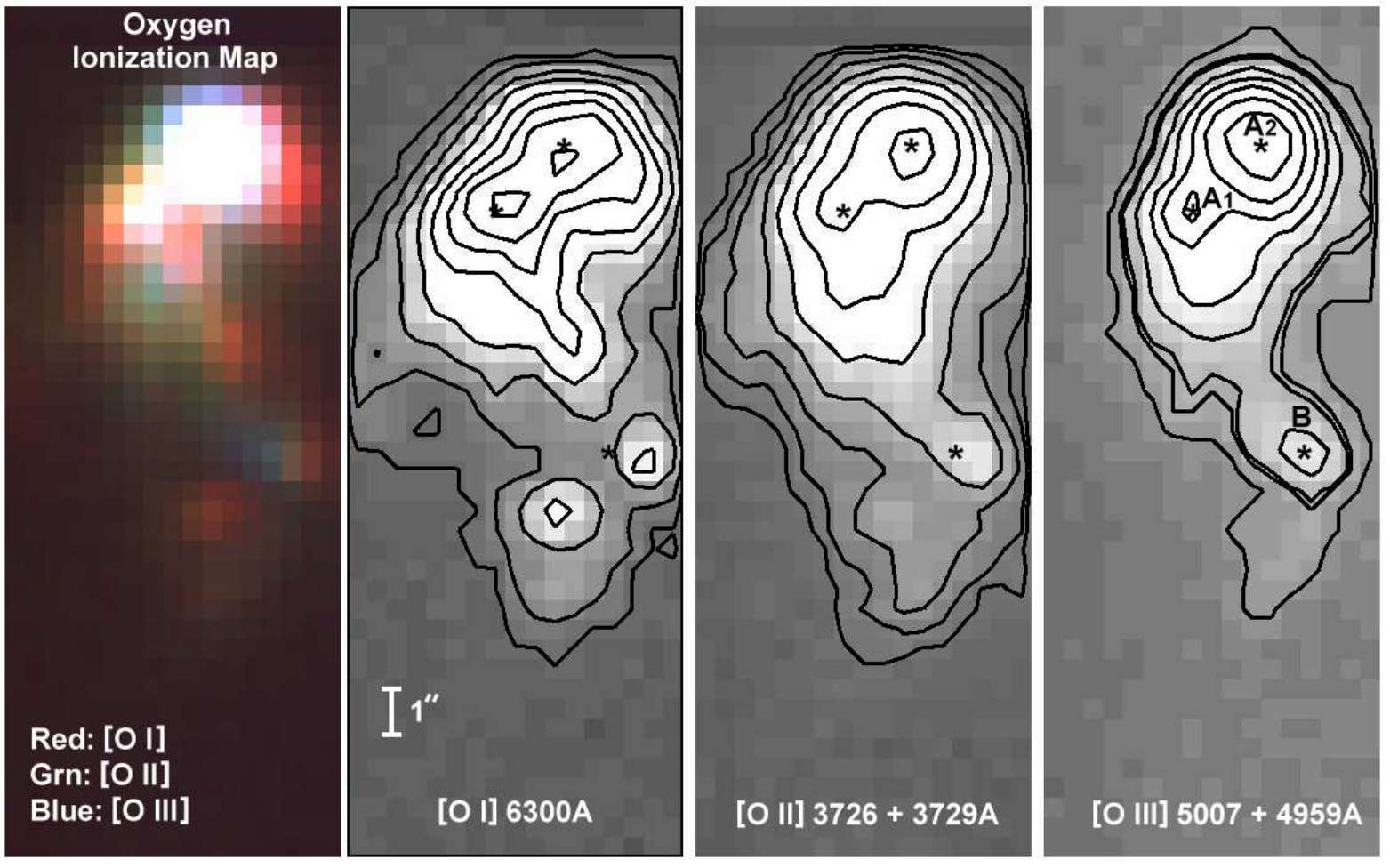}
\caption{Oxygen ionization composite. Contours are drawn in factors of
$\sqrt{3}$ in surface brightness. Each pixel is 0.437\arcsec. The asterisks
denote the peak intensities of knots A1, A2 and B in the [O~III] 5007+4959 image,
and are useful for identifying excitation gradients.
}
\label{fig:oxygen}
\end{figure}

Our data are particularly good for delineating the ionization structure in the
flow because we simultaneously acquired deep data cubes of emission lines from
three ionization states of oxygen, [O~I] $\lambda$6300,
[O~II] $\lambda\lambda$3726+3729, and [O~III] $\lambda\lambda$5007+4959.
We present these cubes side-by-side in Fig.~\ref{fig:oxygen}. Several trends are
worth noting here. First, the [O~I] emission generally covers a broader area
perpendicular to the jet, as expected for a flow that moves nearly along our
line of sight \citep[e.g.][]{beck04}. Second, the A5 area in the middle of the
frame (cf. the middle panel of Fig.~\ref{fig:hstdiff}) emits strongly only
in [O~I], consistent with it being bright in the [S~II] HST image, where it appears
as a wiggling jet. Finally, there is a strong decrease in the ionization from
left to right in knot B as one moves from the center of the jet to
closer to the edge of the flow, as highlighted
by the positional offset between  the asterisk that marks
the [O~III] peak from the peak contours in the [O~I] image. Analogous offsets in knots
A1 and A2 are small. There is also a low-ionization knot about 2\arcsec\ 
to the left of knot B in Fig.~\ref{fig:oxygen}, but this knot and the emission
below it have very high radial velocities of $\gtrsim$ 300\kms\
(Fig.~\ref{fig:cube-intermediate-low}).

\begin{figure}
\centering
\includegraphics[angle=0,scale=1.00]{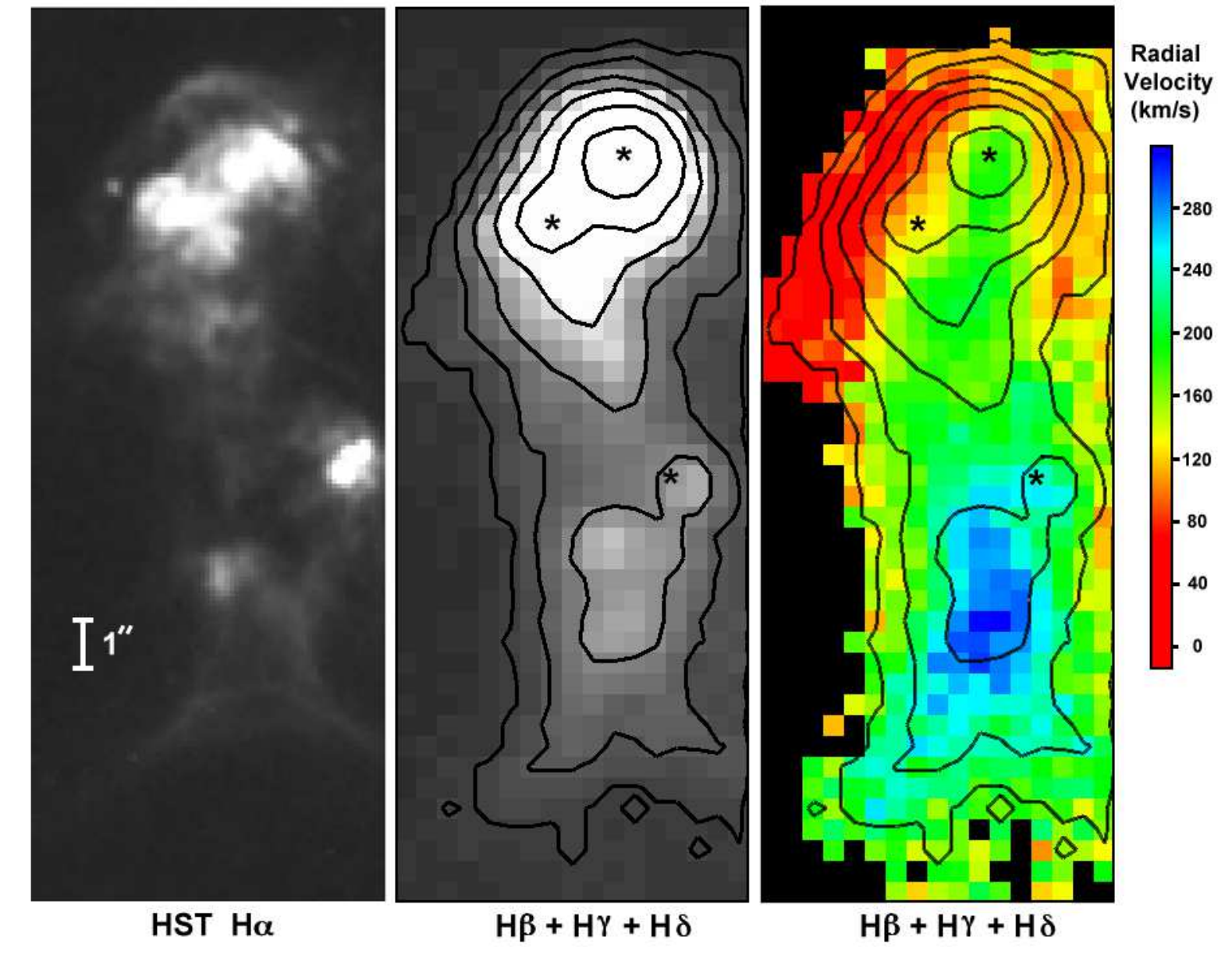}
\caption{Left: HST image at H$\alpha$. Center: Combined H$\beta$ + H$\gamma$ + H$\delta$
image. Adjacent contours differ by a factor of $\sqrt{3}$ in flux.
Right: Average velocity relative to AS~353A in the summed H$\beta$ + H$\gamma$ + H$\delta$ composite.
Asterisks denote the peak emission of knots A1, A2, and B in Fig.~\ref{fig:oxygen}.
}
\label{fig:HIvel}
\end{figure}

Fig.~\ref{fig:HIvel} maps the average Balmer line radial velocities throughout the
region by integrating over the line profile shape at each point
(the equivalent of a moment-one map in molecular data). There is a broad area of
low radial velocity gas that surrounds the outflow, with the fastest material
located along the axis of the flow. Such behavior has been seen before in this
jet \citep{beck04}, and is consistent with a series of bow shocks viewed nearly
along the line of sight.

\begin{figure}
\centering
\includegraphics[angle=0,scale=0.50]{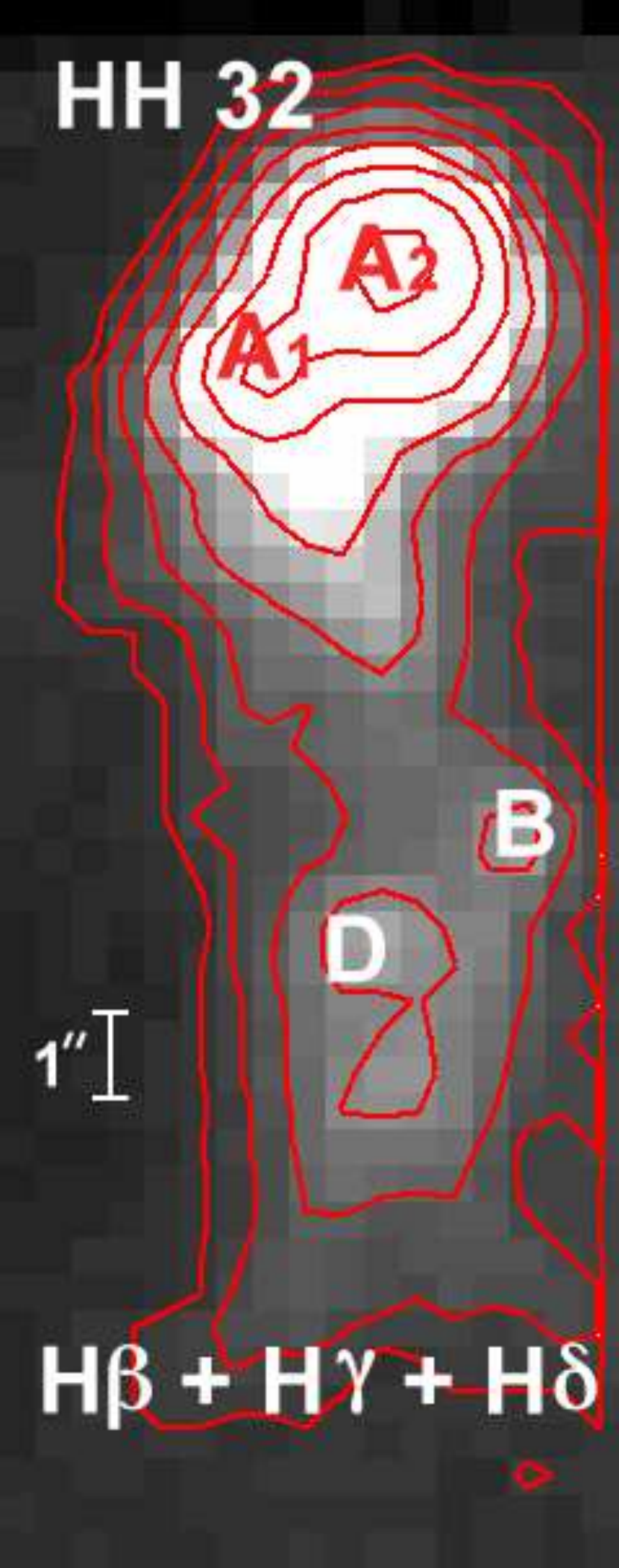}
\caption{Animation through the velocity frames of the KCWI datacube 
at H$\beta$ + H$\gamma$ + H$\delta$.  The animation moves
through each frame between $-$60\kms\ through +450\kms\ in 15\kms\
intervals.  The greyscale placeholder image is integrated over the the full
velocity range, and has contours spaced by a factor of $\sqrt{3}$ in flux
that provide a reference throughout the animation.  The video duration is 12 seconds.
\label{fig:movie}}
\end{figure}

Animations provide a much better way to follow the complex
velocity structure in this outflow (Fig.~\ref{fig:movie}).
We created a movie of the H~I datacube for the on-line version of the paper
that cycles through the radial velocities between $-$60\kms\ to +450\kms\ in 15\kms\
intervals. Moving from blue to red, significant
emission first begins around $-$30\kms\ with a curved filament of
emission near the apex of the bow shock A3 (refer to Fig.~\ref{fig:hstdiff}
for knot names and orientations). By 0\kms\ this emission intensifies to form
a flattened knot above (downstream from) the location of the peak of A2, with a 
wing to the emission to the left ahead of A1.  The right side of knot B
also appears, with a weak extension into the A5 region between B and A1.
Between 0\kms\ and 60\kms, the emission associated with knot A3 shifts down
(upwind) and gets somewhat narrower. At this point the emission is a bright
crescent that aligns with the velocity-integrated
peaks of A1 and A2. Between 60\kms\ and 120\kms\
knot A has begun to split into its two distinct components, A1 and 
A2.  A5 continues to brighen, and by 105\kms\
knot D first appears just as the emission shifts from the right side of knot
B to its left side. Knots D and B now bracket the sides of the jet.

As we continue the journey through the cube, we move from the
low-velocity gas to the intermediate-velocity gas in Fig.~\ref{fig:cube-intermediate-low}.
By 180\kms, emission from both knots B and D have shifted more towards
the axis of the flow, and the `Arc' feature labeled in Fig.~\ref{fig:hstdiff}
at the bottom of the cube has begun to appear. This feature emits in
Balmer lines, but not in the forbidden lines. Emission from the A5 region
has now shifted to the left, and is located directly upstream of A1. Both 
A1 and A2 knots are distinct peaks, and align on the maximum contours
of the integrated cube. Between 180\kms\ and 240\kms\ the emission from
knots B and D merges into a single feature down the axis of the flow,
and the `Arc' at the bottom of the cube nearest to AS 353A
is clearly visible. Knot A1 has all but disappeared, but 
knot A2 remains bright. The A5 emission shifts back to the right, aligned
with the center of the flow. At the high end of the intermediate velocities,
240\kms\ to 300\kms, the merging of knots B and D produces a bright knot
at 300\kms\ along the axis of the flow that is visible both in the contours
of the integrated intensity and in the velocity map (Fig.~\ref{fig:HIvel}).
A faint bridge now connects this emission to a knot in A5, connecting
in turn to A2, where its peak has shifted down (upwind) of the integrated
contour peak by about 0.6\arcsec.

\begin{figure}
\centering
\includegraphics[angle=0,scale=1.00]{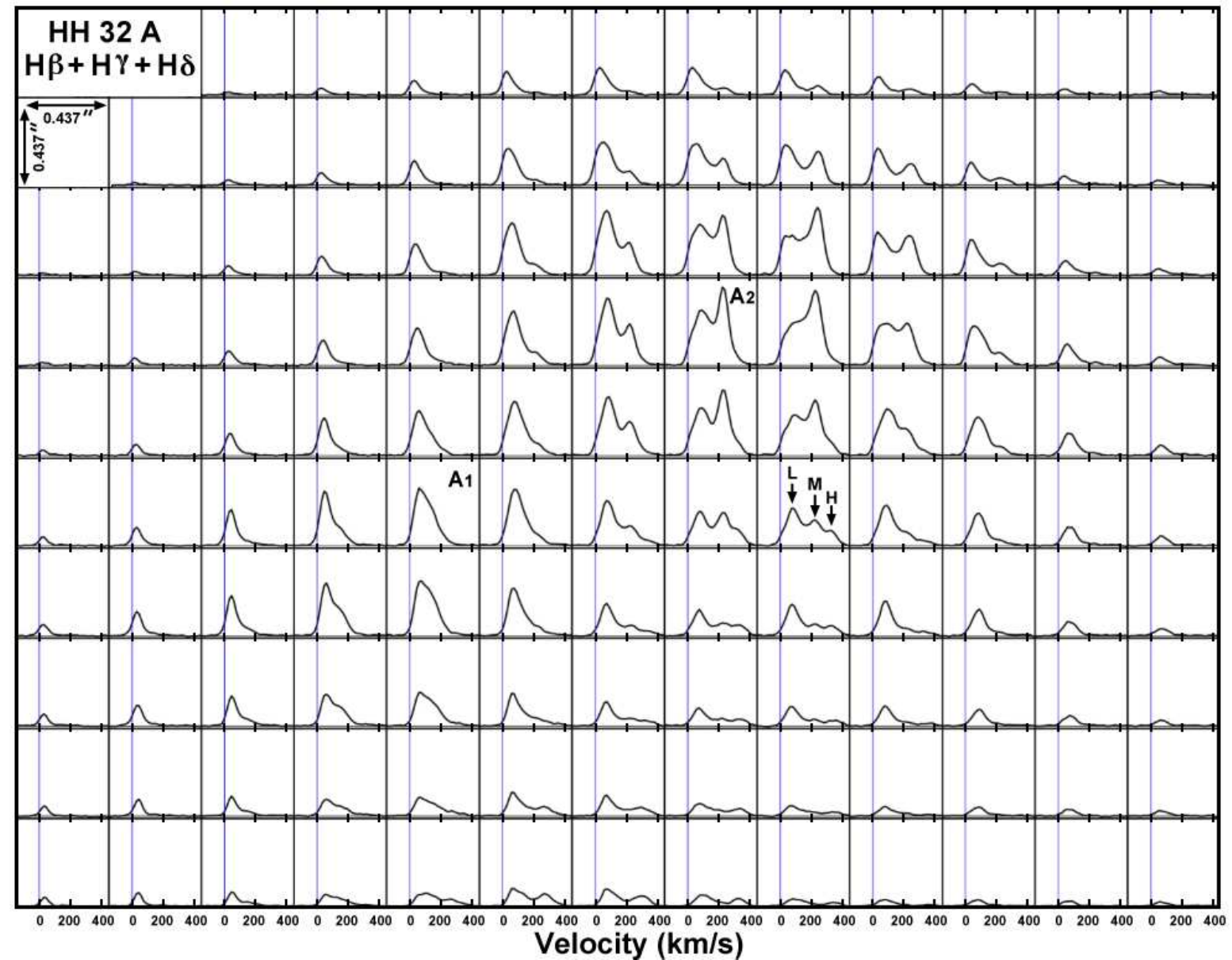}
\caption{Spectral map across HH~32 in H$\beta$ + H$\gamma$ + H$\delta$. Labels mark the
locations of the knots A1 and A2 as defined in the [O~III] image (Fig.~\ref{fig:oxygen}).
The three distinct velocity components labeled `L', `M', and `H' merge into two components
at knot A2. The low-velocity component surrounds the entirety of the emission.
}
\label{fig:hh32a-HI}
\end{figure}

\begin{figure}
\centering
\includegraphics[angle=0,scale=1.00]{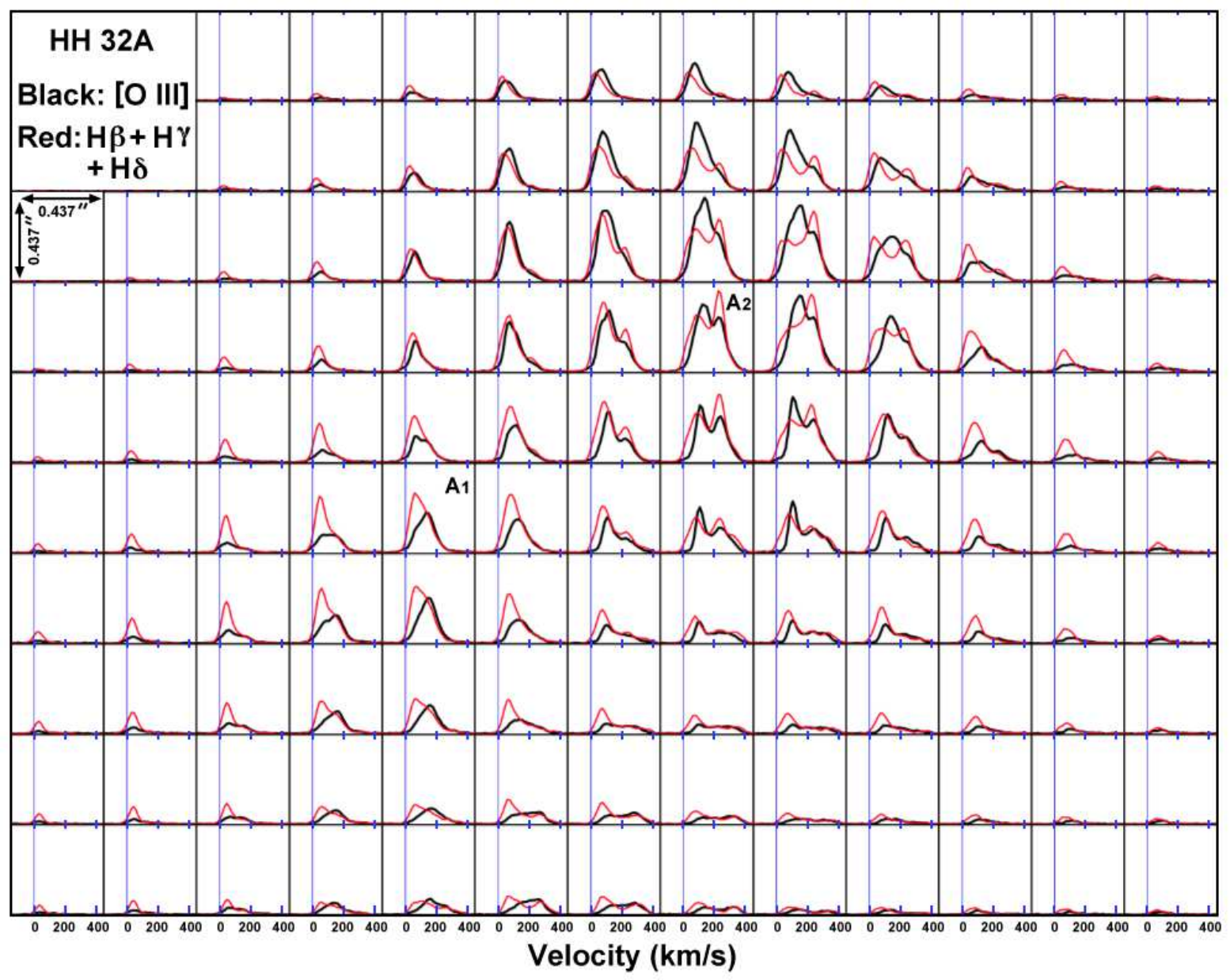}
\caption{Same as Fig.~\ref{fig:hh32a-HI} but superposing the [O~III] line
profiles on top of the H~I line profiles.
}
\label{fig:hh32a-oiii-hI}
\end{figure}

\begin{figure}
\centering
\includegraphics[angle=0,scale=1.00]{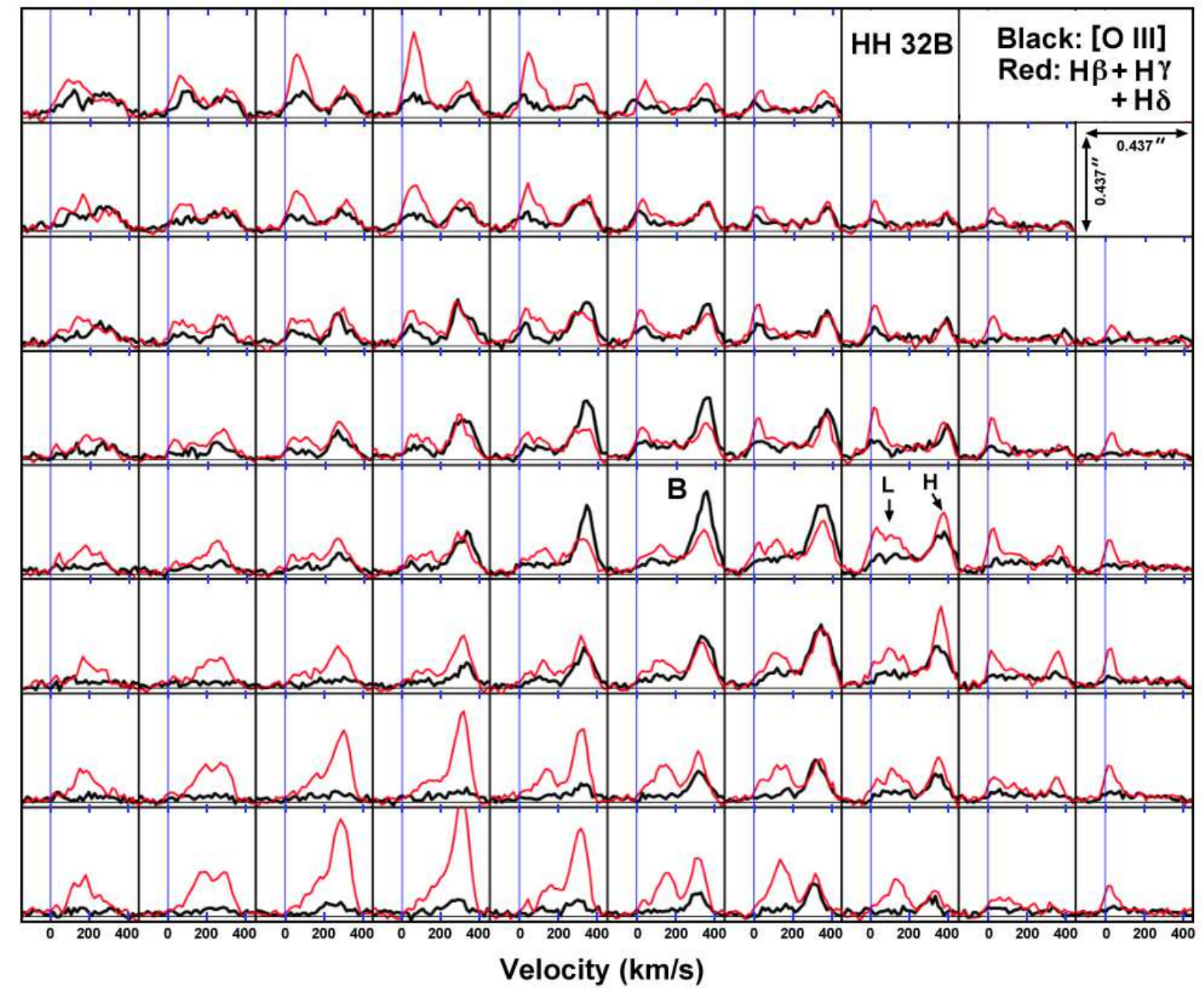}
\caption{Same as Fig.~\ref{fig:hh32a-HI} but for the area around knot B. The location
of knot B, defined by the peak of the [O~III] emission, is marked.
}
\label{fig:hh32b-oiii-hI}
\end{figure}

Between 300\kms\ and 360\kms\ the knots of A5 and A2 merge into
a single linear feature centered on the jet's axis and located upstream from the
peak emission of knot A2. Emission along the axis between knots B and
D remains strong. The north side of knot B reappears as an emission source around 360\kms.
Above 360\kms, emission along the jet gradually fades, and disappears by
420\kms. Knot B remains visible to slightly higher velocities, and all
emission is gone by 450\kms.  The full complexity of the outflow is on display in Figs.~\ref{fig:hh32a-HI}
through \ref{fig:hh32b-oiii-hI}, where we plot the spectra of various lines
for each pixel.  Fig.~\ref{fig:hh32b-oiii-hI} presents
a similar spectral map for the region around knot B.
We discuss these maps more in the next section.

\clearpage

\section{Discussion} \label{sec:discussion}

\begin{figure}
\centering
\includegraphics[angle=0,scale=1.10]{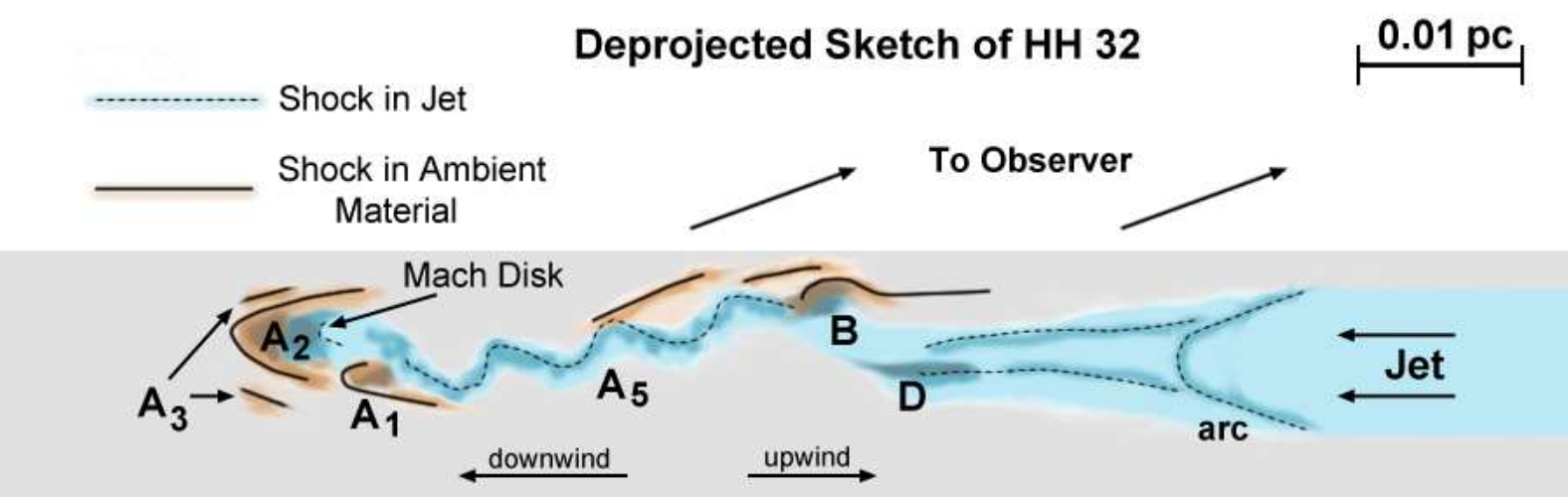}
\caption{
A heuristic model for the HH 32 system, viewed perpendicular to the flow direction.
The position of the features derives from deprojecting the HST images using an
observing angle of 20 degrees to the line of sight. Blue colors indicate jet material
and orange colors indicate ambient gas, with the shock waves in those materials
depicted by dashed lines
and solid lines, respectively. The major features of the jet are labeled. The jet
flows from right to left. The driving source AS~353A would be located well off the frame
($\sim$ 0.045~pc) to the right of this diagram (Fig.~\ref{fig:overview}.)
}
\label{fig:cartoon}
\end{figure}

Even though the spatial and velocity structures within HH~32 are complex,
we can understand most of what we observe in the HST images and in the KCWI
datacubes by using the standard picture of a highly
supersonic jet with a variable direction and speed. Fig.~\ref{fig:cartoon}
depicts how the jet should appear if it were viewed perpendicular to the direction
of the flow. To construct this figure to scale, we 
stretched the HST image along the axis of the jet to account for a
viewing angle of 20 degrees to the line of sight as inferred from proper
motion measurements \citep[e.g., Fig.~10 of][]{curiel97}.  Fig.~\ref{fig:cartoon}
labels the major features along the flow, indicates
the shapes of the main shock waves, depicts whether or
not the shocks occur in ambient (or very slowly moving) gas or in the
jet, and outlines an approximate boundary between the jet flow and the surrounding
medium.

The following section describes the observational evidence to support
the scenario drawn for each of the features in Fig.~\ref{fig:cartoon}, keeping
in mind the inherent limitations imposed by the lack of axial symmetry
in the flow. The location of the 
boundary between the jet and ambient material is often uncertain, as the gas
typically only becomes visible as it passes through a shock front, so not
all portions of the flow are traced by the emission line observations. 

\subsection{Schematic for the HH~32 Redshifted Jet}

{\it The Arc - Weak Shocks in the Jet}:

The structure labeled `Arc' in Fig.~\ref{fig:hstdiff} is the closest
emission line feature to AS~353A in the redshifted jet. The HST
images show only H$\alpha$ and no [S~II] at this location, though the Arc
is visible faintly in the HST narrowband [N~II] $\lambda$6583 image of \citet{curiel97}.
The Arc appears only in our Balmer composites with KCWI and not in [N~II] $\lambda$5755 line
or in any of the other forbidden lines, though [N~II] $\lambda$5755
is one of the fainter lines in our cubes (Fig.~\ref{fig:cube-intermediate-hi}).
The radial velocity in the Balmer lines is high in Fig.~\ref{fig:cube-intermediate-low},
about 250\kms.  The emission line morphologies and velocities all imply that the Arc is
a weak shock in the jet. Linewidths are narrow, there are no bright knots to define
a cooling zone, and radial velocities are high. 

The simplest explanation for the Arc is that the feature forms as a pulse of faster material
overtakes slower material in the jet. In this scenario, the Arc represents a weak
bow shock which forms as 250\kms\ pulse overtakes somewhat slower gas ahead of it. 
This model predicts the Arc will have a high proper motion, similar to the other
features in the redshifted jet. 

If the Arc has no proper motion, it probably represents a focusing shock.
These oblique shocks occur as the jet responds to a sudden increase in the ambient density,
or if the jet `overexpands' and the ambient medium redirects the jet back towards its axis.
The concept is similar to the focusing-cavity jet model of \cite{canto80},
with the important difference that the external medium
collimates a spherical flow from the source in the Cant\'o model, whereas here the flow must
already be collimated when it encounters the Arc because the Arc subtends a small
solid angle as viewed from the driving source AS~353A. The continuum HST image (Fig.~\ref{fig:overview})
shows that the AS 353 binary is embedded within a dark cloud and appears to have
evacuated a cavity in the cloud. With this geometry it is plausible that a jet
would burrow through a cloud along its path, and focusing shocks would mark the location
where the jet enters the cloud. However, this scenario only works if the Arc has
no proper motion.

{\it Knot D - Another Jet Pulse}:

In the HST images, knot D is a bright linear feature situated a bit to the left of the axis of the 
flow in Figs.~\ref{fig:hstdiff} and~\ref{fig:cartoon}. A faint filament of H$\alpha$ emission precedes knot
D along the the jet. The KCWI composites look somewhat different, with, for example, the
integrated [O~I] image showing a knot aligned to the axis of the jet (Fig.~\ref{fig:oxygen}).
Taken at face value, it means that knot D has formed forbidden lines at its apex
in the 25 years since the Hubble images. The datacubes show an increase in velocity along
the center of the jet (Sec.~\ref{sec:vel}). All of these observations are consistent with
a jet pulse of $\sim$ 300\kms\ that overtakes material in front of it. The velocity difference
between this pulse and the one in front of it must be $\lesssim$ 100\kms\ to account for the
lack of [O~III] emission in Fig.~\ref{fig:oxygen}.

{\it Knot B - Partial Bow Shock into a Cavity Wall}:

Together with knots A1 and A2, knot B is one of the brightest features in HH 32
and has plenty of [O~III] emission. 
Hence, it must represent a strong shock wave. The HST image shows two or three
compact knots, depending on the emission line. The knot is located on the southern
boundary of the jet (Fig.~\ref{fig:overview}), and as noted in Sec.~\ref{sec:vel} and
Fig.~\ref{fig:oxygen}, there are strong ionization and velocity gradients in this 
area, with both velocity and excitation increasing towards the axis of the jet.
The r-band HST image (Fig.~\ref{fig:overview}) shows what appears to be a spur shock
into the surrounding gas, and the velocity of that material is low, a few tens of \kms\
(Figs.~\ref{fig:HIvel} and~\ref{fig:hh32b-oiii-hI}, Sec.~\ref{sec:vel}). 

Knot B has all the hallmarks of a partial bow shock moving into ambient material. Its
location along the edge of the flow means a shock will encounter ambient gas there, and
the wing of the bow shock becomes the spur shock into the ambient gas. Weak shocks in
the ambient molecular gas then produce the H$_2$ emission observed along the interface
between the jet and the ambient gas by \citet{davis96}.  Spectral
maps of the knot B region show distinct low-velocity and high-velocity
components (labeled `L' and `H', respectively in Fig.~\ref{fig:hh32b-oiii-hI}) over
the entire area. The low-velocity material extends to the side of the jet beyond
the high-velocity material. [O~III] is brightest in the high-velocity gas
where its emission peaks at the location of knot B. In a bow shock model, this emission
arises from the apex of the bow ($\sim$ 330\kms), while the Balmer line emission
distributes more evenly across the bow and results in more of a double-peaked emission
line profile \citep[e.g.][]{sbr86,hrh87}. 
The northern side of the bow (left side in Figs.~\ref{fig:hstdiff}, \ref{fig:HIvel} and
\ref{fig:hh32b-oiii-hI} and down in Fig.~\ref{fig:cartoon}), is
replaced by weaker shocks which serve to deflect the flow away from the cavity. These
shocks produce emission only at high velocities, as is observed. In fact, one can
connect the fastest portion of the jet in [O~I] $\lambda$6300 from the northern side of knot B all the way
to the Mach disk in knot A2 (Fig.~\ref{fig:cube-intermediate-low}).

The partial bow shock model described above does not explain the multiple knots in the
HST images. Shock velocities over 200\kms\ like those present in knot B are
prone to cooling instabilities that could form knots \citep[e.g.][]{suzuki15}, or the flow may simply
be clumpy on size scales of a few hundred AU. Ground-based datacubes such as ours
do not have enough spatial resolution to probe the velocity structures of features
on these size scales.

{\it A5 - Wiggling Jet Projected Along the Line of Sight}:

Region A5 connects knot B to knots A1 and A2. The [S~II]
HST image has a remarkable morphology of 3 $-$ 4 sharp bends
that resemble a rope dropped onto the floor.  Velocities here are
among the highest in the jet (Sec.~\ref{sec:vel}).
The sinuous feature emits in [O~I], [O~II], and [O~III]. 
It is difficult to trace exactly where the jet goes near
A1. The velocity images (Figs.~\ref{fig:cube-hi} $-$ \ref{fig:cube-low})
show the highest velocity material turns back to the south near A1 and
connects seamlessly with knot A2. The H$\alpha$ datacubes of \citet{beck04}
enjoyed 0.5\arcsec\ seeing, and
show a separate faint knot at 395\kms\ with weak bow-like tails on
either side at this location. Our Balmer slice also shows this feature, though
we are unable to make out the wings. This is the region where the jet feeds
into the main bow shock at A2.

We identify the sinuous feature with a wiggling jet. Such structures are common
in HH flows, and typically emit only in low-excitation lines as they are
excited by weak internal shocks.
Deprojecting the flow shows that one can fit $\sim$ 4 sinusoidal variations
with modest amplitudes in a jet between knot B and knot A1 and still have that
portion of the jet project onto one small area of the sky (Fig.~\ref{fig:cartoon}).
Hence, the severe bends present in the image of the A5 jet result from
a projection angle nearly along the line of sight.

Overall, the kinematics and morphology of A5 are well-explained by a 
simple wiggling jet viewed nearly along the line of sight. The strong [O~III]
in this region is unusual however (Fig.~\ref{fig:cube-hi}). To understand why this occurs
we would need an HST-resolution image at [O~III] and attempt to analyze the shock waves at
each position along the jet. One other unusual aspect of the A5 region
is the presence of extensive low-velocity gas in the spectral maps, especially in the Balmer
lines (component `L' in Figs.~\ref{fig:hh32b-oiii-hI} and \ref{fig:hh32a-HI}).
The HST difference image in Fig.~\ref{fig:hstdiff} hints as to what this emission might be: 
a Balmer-only filament exists downwind from knot B and connects to the wiggling
jet in A5. This feature is likely to be a weak shock that propagates into
the ambient material to provide a `sheath' of low-velocity Balmer emission that
surrounds the jet (the `spur' shock drawn in region A5 in  Fig.~\ref{fig:cartoon}). 
The faint H$_2$ emission in this area can also arise from these weak shocks \citep{davis96}.

{\it Knot A1 - Classic Bow Shock}:

\citet{beck04} found they were able to explain the velocity and spatial
structure of the bright knot A1 in their H$\alpha$ cubes
remarkably well with a simple bow shock model. Our new datacubes
fully support their interpretation. The spectral maps of A1 have an extended 
low-velocity halo that surrounds the high-velocity peak, and the
[O~III] emission peaks at a higher radial velocity than the Balmer lines do.
All of these features agree with the predictions of bow shocks viewed
nearly along the line of sight \citep{beck04,raga04}. 

The region is rather complex in the HST images. As noted above in the discussion of A5, a fast jet knot
is superposed about 0.5\arcsec\ to the south of the brightest part of A1, presumably
on its way to A2. The bright knot in A1 is in the correct location to be a Mach
disk, but it is hard to know for certain because 
the apex of the bow shock also projects to this location at this orientation, and the rest
of the region, including the A1 bow shock, is quite clumpy on small spatial scales.
The [O~III] emission peaks at around 150\kms\ in knot A1 (Fig.~\ref{fig:hh32a-oiii-hI}),
so this is what the velocity of the working surface should be, and will also be the
shock velocity because this knot appears to move into ambient gas. 
Bow shocks need not have Mach disks for episodic flows, because if the jet shuts off
the Mach disk will vanish while the bow shock continues to propagate.

{\it Knot A2 - Bow Shock with a Mach Disk}:

Knot A2 shares many of the same characteristics as knot A1, including a
bright, compact, high-velocity and high-excitation core surrounded by a low-velocity
halo. Fig.~\ref{fig:hstdiff} shows A2 has an extended bow shock, but the brightest
knot in the region is very compact, and not clearly resolved even with HST. 
Both the location of this knot relative to the putative bow shock and the fact that it
aligns with the fastest part of the jet in A5 support its identification as a Mach disk. 

The kinematics of the flow also support a Mach disk interpretation for the
object. Fig.~\ref{fig:hh32a-HI} shows three distinct velocity components 
in the Balmer lines labeled L, M, and H, where component L surrounds the entire flow
as a low-velocity sheath. As the jet moves downwind into knot A2, component
H at $\sim$ 325\kms\ vanishes and the flux in component M at $\sim$ 230\kms\
suddenly rises. We interpret this 230\kms\ value as the velocity of the working
surface, in which case the Mach disk would have a shock velocity of $\sim$ 100\kms. 

{\it Knots A3 and A4 - Possible Lateral Radiative Precursor}:

Feature A3 actually consists of two arcs, one located just beyond
A1 and another associated with A2. Along the southern edge of
the flow, A3 continues with a feature known as A4 (Fig.~\ref{fig:hstdiff}).
Although A3 and A4
seem to lie downwind of knots A1 and A2, Fig.~\ref{fig:cartoon} shows that
this situation is not necessarily the case for a viewing angle near to the line of sight.
The radial velocities in A3 and A4 are near zero (Fig.~\ref{fig:HIvel}), and
in fact these areas have a bit of blueshifted emission, and appear in the $-$30\kms\
slices.  Linewidths in this region are typically $\lesssim$ 50\kms. These regions also
have narrow H$_2$ emission at the ambient velocity, though the spatial resolution of
the H$_2$ observations makes it difficult to tell if the emission comes from A3 and
A4 or the downwind edges of A1 and A2 \citep{davis96}.

A radiative or magnetic precursor is an attractive model for A3 and A4 because one
would expect the precursor to follow the overall outline of the main shock, but retain
a linewidth close to the thermal speed and have little radial motion. For the
orientation of HH~32, the precursor could lie along the side of the jet and
still appear to be projected ahead of it. HH~32 is one of the few HH outflows
with high enough shock velocities to expect to see a radiative precursor. Effectively
the shock front creates a small H~II region that follows it along on its journey.
This phenomenon has been observed in the bright HH knot HH~2A', which also emits
strongly in [O~III] \citep{hartigan11}. If the precursor idea is correct, A3 and
A4 should have the same proper motions as the high-excitation knots A1 and
A2. If the radiative precursor is bright enough, it
will cause fluorescence in H$_2$ that may be observable at 2.12$\mu$m..

It is also possible that A3 and A4 represent the edges of a cavity evacuated
by a previous ejection, and a weak shock propagates into the ambient gas at these
locations.  Though there is no evidence for a larger scale outflow,
if we take the velocity of the bow shock A2 to be 230\kms\ and the deprojected
distance of knot A2 from AS~353A to be 0.12~pc (Fig.~\ref{fig:overview}),
the travel time of knot A2 to its current location is only about 500 years,
3 $-$ 4 orders of magnitude shorter than the likely age of the star. Hence,
knot A2 is almost certainly not the first ejection from the system, though it could
still propagate into ambient gas if the outflow is sufficiently time variable
that the ambient medium refills the jet cavity between major outbursts.

\begin{figure}
\centering
\includegraphics[angle=0,scale=1.20]{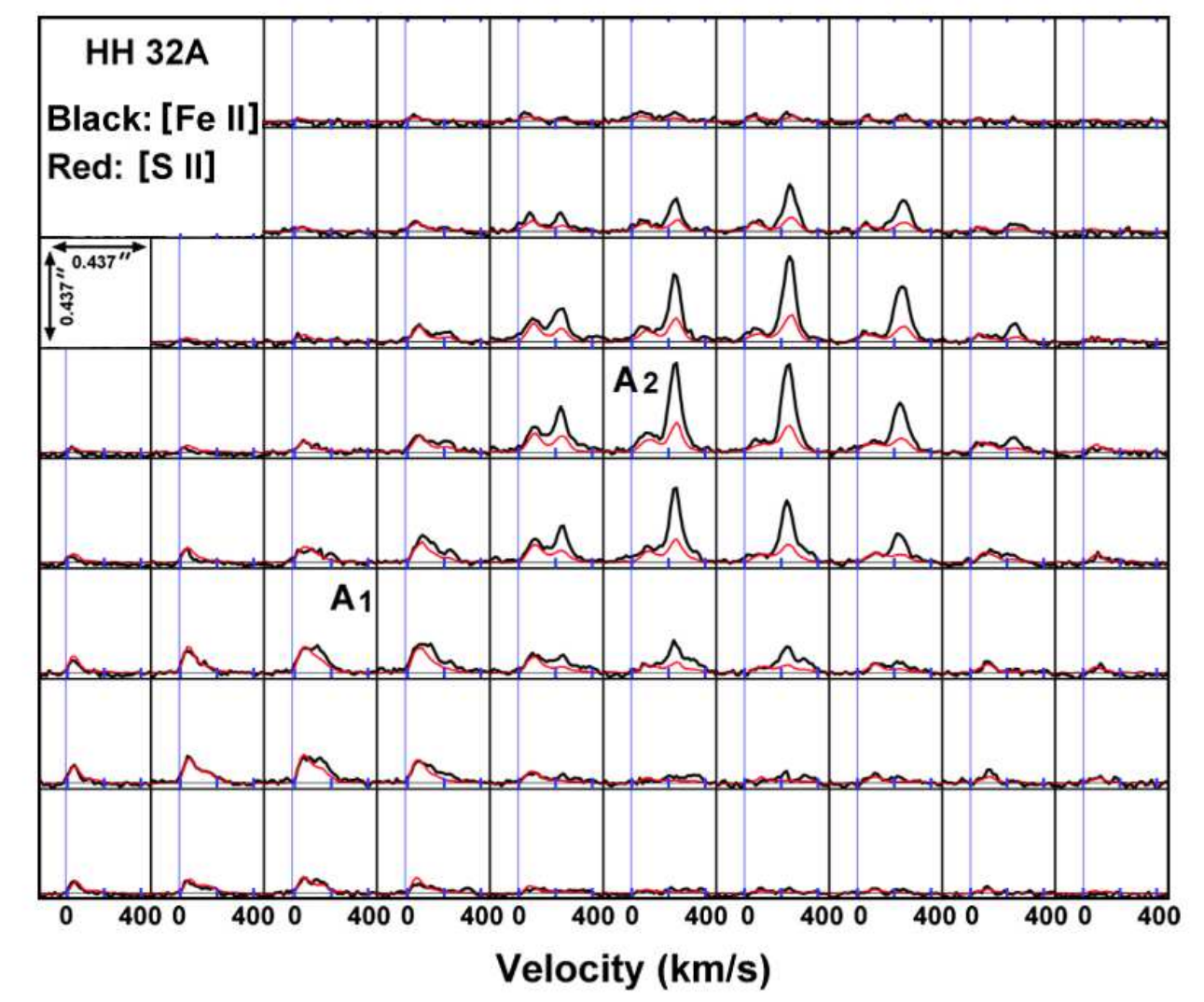}
\caption{Spectral map of the knot A region, superposing the [Fe~II] line
profiles on top of the [S~II] line profiles. These lines should track one-another
closely; however, the data show a marked increase in one-velocity component
in the [Fe II] lines at knot A2 which we interpret as evidence for dust
destruction, where the shock releases refractory elements into the gas phase.
}
\label{fig:depletion-fe}
\end{figure}

\begin{figure}
\centering
\includegraphics[angle=0,scale=1.20]{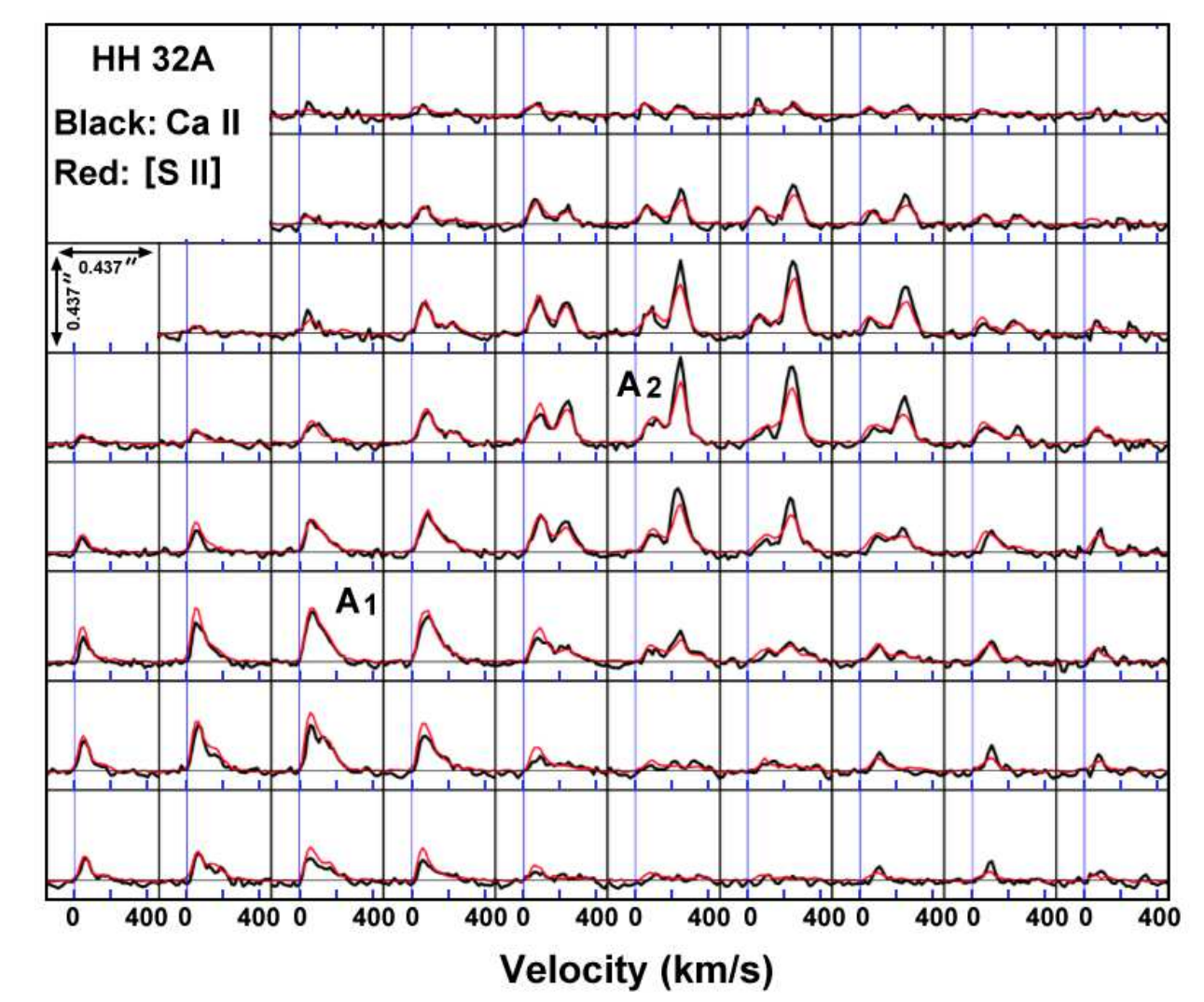}
\caption{Same as Fig.~\ref{fig:depletion-fe} but for Ca~II $\lambda$3933. There are no
systematic differences between the [S~II] and Ca~II line profiles. 
}
\label{fig:depletion-ca}
\end{figure}

\begin{figure}
\centering
\includegraphics[angle=0,scale=1.20]{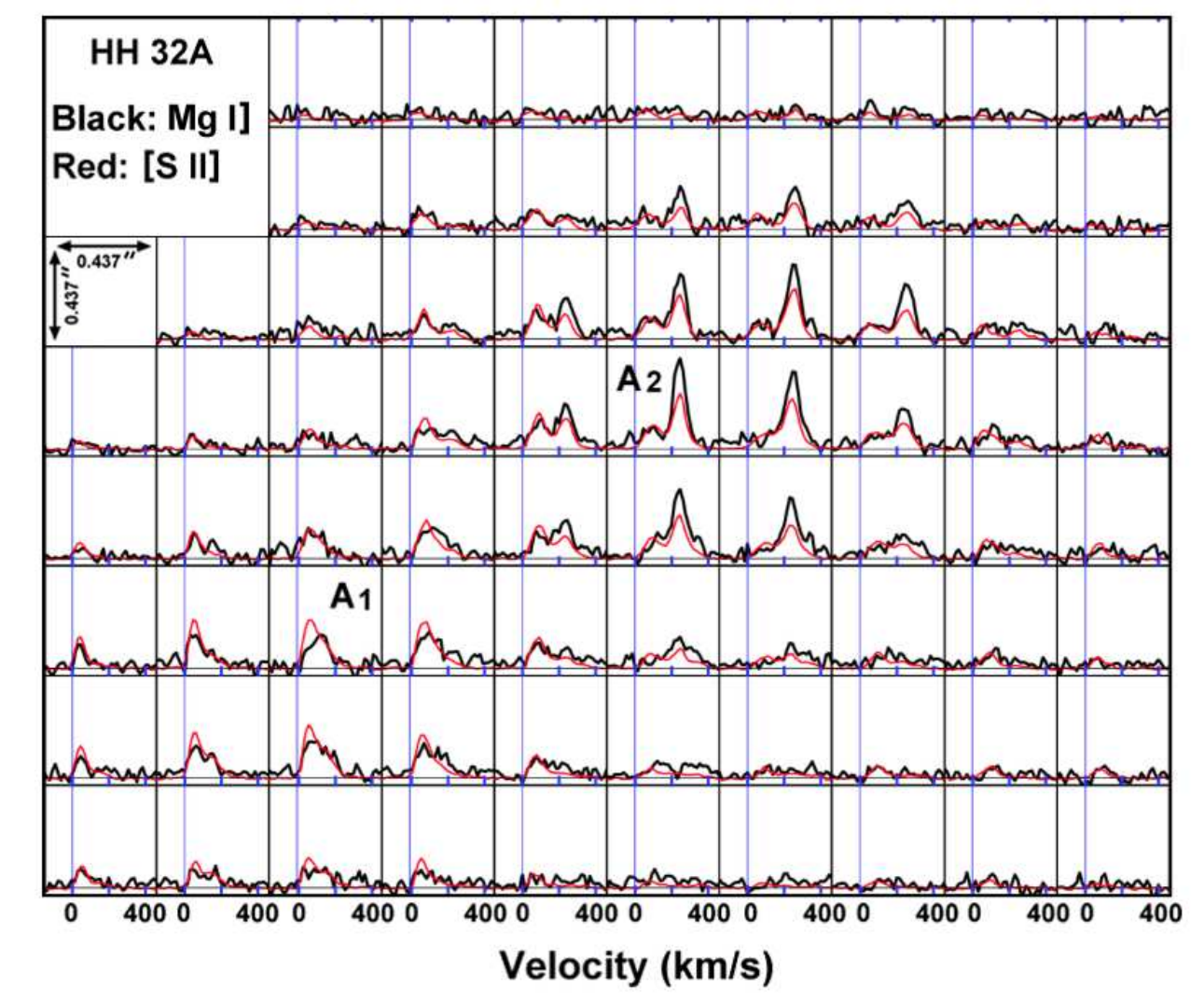}
\caption{Same as Fig.~\ref{fig:depletion-fe} but for the Mg I] 4571 line.
}
\label{fig:depletion-mg}
\end{figure}

Overall, the simple scenarios outlined above are remarkably successful in 
explaining most of the complexities in our data cubes. However, the HST images
remind us to not be overconfident, as there are many overlapping knots and
filaments that become resolved on subarcsecond scales.  To address that
level of detail scientifically would require combining high-dispersion spectra 
on these spatial scales with temporal observations that can distinguish
internal proper motion differences. Such observations are beyond what we can
do in the present work, and could be difficult to interpret owing to the
orientation of this outflow, which essentially projects the entire jet onto
a small area.

\subsection{Spectroscopic Evidence for Dust Destruction}

The spectral maps of the low-excitation lines ([Fe~II] and [S~II] composites, Ca~II $\lambda$3933
and Mg~I] $\lambda$4571) in the region of knot A1 and A2 in Figs.~\ref{fig:depletion-fe},
\ref{fig:depletion-ca}, and~\ref{fig:depletion-mg} are generally quite similar. However,
one discrepancy stands out sharply:
the velocity component we identify as belonging to the Mach disk of knot A2 has a 
larger [Fe~II]/[S~II] ratio by a factor of two relative to that in
the low-velocity component and throughout the profiles in the rest of the region. 
This enhancement is not present in the Ca~II/[S~II] ratio, and appears only
weakly in Mg~I]/[S~II]. 

The easiest way to explain the enhanced Fe~II flux in knot A2
is if the shock waves there convert iron dust grains into gas.
Dust grains can be destroyed in shocks both by grain-grain collisions
and through sputtering by atoms and ions \citep[see][for a review]{jones04}. Shattering
by grain-grain collisions breaks large grains into smaller pieces
for impact velocities as low as 1 km$\,$s$^{-1}$ \citep{jones96},
but this process simply changes the size distribution of the grains without
modifying the gas to dust mass ratio unless the impact velocities
exceed the vaporization threshold of $\sim$ 20 km$\,$s$^{-1}$ \citep{tielens94}.
Above about 50 km$\,$s$^{-1}$, sputtering dominates
dust destruction and converts $\sim$ 50\%\ of the mass of an
iron grain into gas for injection velocities $\gtrsim$ 170 km$\,$s$^{-1}$
\citep[Fig.~11 of ][]{jones96}, and a similar fraction for C and Si grains
\citep[Fig.~7 of ][]{slavin15}.  In the case of knot A2, the shock wave is 
a fairly strong one, with a shock velocity $\sim$ 100\kms\ if it is the Mach disk and
$\sim$ 230\kms\ if it is the apex of the bow shock. In either case such a
strong shock could substantially increase the gas phase abundance of iron
at this location.  Iron is highly refractory, and some evidence exists for its depletion in jets
\citep{dust14}. Our data seem to be the first direct evidence for a refractory element being returned
to the gas phase by a shock in a stellar jet, however.

\section{Conclusions} \label{sec:conclusions}

In this paper we have acquired and analyzed over 60 datacubes of the 
stellar jet HH~32. The cubes have $\sim$ 30\kms\ velocity resolution
and $\lesssim$ 1\arcsec\ seeing. The project focused on blue lines,
which enables the first detailed study of several new ionization states of 
common elements in stellar jets, including datacubes of Ca~II, Mg~I, O~I,
O~II, O~III, He~I, He~II, Fe~III, Fe~II, S~II, N~II, and three Balmer lines.
We found that the overall morphologies of the line emission within these cubes
sort remarkably well when grouped according to excitation, defined
as a combination of the ionization potential of the element and the
energy level of the upper state above ground.  

The results generally agree with previous works in the areas of overlap, and
fit well with a scenario of a pulsed jet with variable ejection angle that we
view nearly along line of sight.  Knots A1 and A2 exhibit
all the kinematic and excitation signatures expected for a resolved bow shock.
Bright areas in these regions are likely to be Mach disks where jet material
enters a working surface. The A1 and A2 bows are fed by A5, a high-velocity wiggling jet.
Closer to the source, knot B appears to be a partial bow shock, while knots
D and a feature known as the Arc are shocks within the jet material.
The spectral cubes of the Mach disk in knot A2 show a sudden jump the
the [Fe~II] flux that we attribute to an increase of iron in the gas phase
as a result of dust destruction in that shock. A similar jump is absent in
Ca~II and only weakly visible in Mg~I]. Dust could have been entrained into
the flow, or even launched in the jet from its circumstellar disk, although in 
these scenarios the dust would have to survive being accelerated to over 300\kms\
without being destroyed.  The extended
low-velocity filaments A3 and A4 that appear ahead of the main bow shocks
could identify a magnetic or radiative precursor to the main shocks, or
may represent the walls of a cavity formed by a previous ejection. 

These observations show the power (and challenges) of combining high-spectral
resolution datacubes of many lines with narrow-band HST images. Without 
spectral information, it is easy to misinterpret HST images owing to
projection effects, even when multiple filters are available. Alternatively,
data cubes taken with ground-based resolution blur the geometry,
leading to simplistic interpretations that change markedly once fragments and filaments
become resolved.  The blue spectral region has the advantage of being able to acquire
velocity cubes in O~I, O~II, and O~III, and also samples both very high-excitation lines
(e.g. He~II $\lambda$4686) and very low-excitation lines (e.g. Mg I] $\lambda$4571). 
A downside is that there are enough lines that blends begin to become
a problem, though one can deblend the [O~II] doublet reliably under
most circumstances.

\acknowledgments
We thank Luca Rizzi for his participation in these KCWI comissioning observations.
This work has made use of data from the European Space Agency (ESA) mission
{\it Gaia} processed by the {\it Gaia} Data Processing and Analysis Consortium.
Funding for the DPAC has been provided by national institutions, in particular the institutions
participating in the {\it Gaia} Multilateral Agreement.

%

\vspace{5mm}
\facilities{Keck (KCWI)}

\end{document}